\documentclass[11pt]{article}

\usepackage{psfrag}
\usepackage{amsthm}
\usepackage{amssymb}
\usepackage{graphicx}
\usepackage{amsmath}
\usepackage{amsfonts}

\usepackage{pstricks}
\psset{unit=1cm}
\psset{arrowsize=1cm}
\psset{linewidth=1cm}

\newtheorem{theorem}{Theorem}
\newtheorem{proposition}[theorem]{Proposition}

\newcommand{\itl}{\textit}
\newcommand{\beg}{\begin}

\newcommand{\non}{\nonumber}
\newcommand{\noi}{\noindent}

\newcommand{\mf}[1]{\mathfrak{#1}}

\parskip=\medskipamount
\title{A Generalization of Chaplygin's Reducibility Theorem}
\author{ O.E.\ Fernandez$^{a}$\footnote{oscarum@umich.edu}\,, T.\
Mestdag$^{b}$\footnote{tom.mestdag@ugent.be} and A.M.\ Bloch$^{a}$\footnote{abloch@umich.edu}\,\\[2mm] {\small $^{a}$Department
of Mathematics, University of Michigan,}\\ {\small 530 Church
Street, Ann
Arbor, MI-48109, USA} \\[2mm] {\small $^{b}$Department of Mathematical Physics and
Astronomy, Ghent University,}\\ {\small Krijgslaan 281, S9, 9000
Gent, Belgium}}
\date{}

\begin{document}

\maketitle


\small
In this paper we study Chaplygin's Reducibility Theorem and extend its applicability to nonholonomic systems with symmetry described by the Hamilton-Poincar\'e-d'Alembert equations in arbitrary degrees of freedom. As special cases we extract the extension of the Theorem to nonholonomic Chaplygin systems with nonabelian symmetry groups as well as Euler-Poincar\'e-Suslov systems in arbitrary degrees of freedom. In the latter case, we also extend the Hamiltonization Theorem to nonholonomic systems which do not possess an invariant measure. Lastly, we extend previous work on conditionally variational systems using the results above. We illustrate the results through various examples of well-known nonholonomic systems.
\large


\section*{Introduction}


Although it is well known that nonholonomic mechanical systems are not variational \cite{B} and thus their mechanics cannot be expressed in terms of canonical Hamilton equations, nevertheless several authors (dating back at least as early as S.A. Chaplygin \cite{Chap,Chap2} and Appell \cite{Ap}) have attempted to express the mechanics of nonholonomic systems in Hamilton-like forms through several methods. Perhaps the most well-known of these methods is Chaplygin's own \itl{Reducibility Theorem}, whose first part states that for nonholonomic systems in two generalized coordinates $(q_{1},q_{2})$ possessing an invariant measure with density $N(q_{1},q_{2})$, the equations of motion can be written in Hamiltonian form after the time reparameterization $d\tau = N dt$ (in this context $N$ is known as the \itl{reducing multiplier}, or simply the \itl{multiplier}). The second part of the Theorem (see \cite{FJ}) says that if a nonholonomic system can be written in Hamiltonian form after the time reparameterization $d\tau=f(q)dt$, then the original system has an invariant measure with density $f^{m-1}(q)$, where $m$ is the degrees of freedom and the function $f$ is again called the {\em reducing multiplier}, or simply the {\em multiplier}. Since both components of the theorem involve a reparameterization of a nonholonomic system into a Hamiltonian one, one often refers to this as the \itl{Hamiltonization} of a nonholonomic system, although we shall refer to it here as the  \itl{Chaplygin Hamiltonization} instead\footnote{We introduce this term because there are other ways of writing the reduced, constrained mechanics of a nonholonomic system as a Hamiltonian system that do not involve a time reparameterization, for example as was done in \cite{BFM,FB}.}.

Chaplygin's original motivation for such a Hamiltonization of nonholonomic systems seems to have been rooted in his interest in the explicit integrability of nonholonomic systems. Indeed,  in \cite{Chap2} Chaplygin applies his method to integrate what would later become known as the Chaplygin sleigh (see Section \ref{CSS}), and remarks that his general procedure (using the reducing multiplier) for integrating certain two degree of freedom nonholonomic systems is ``interesting from a theoretical standpoint as a direct extension of the Jacobi method to simple nonholonomic systems.'' Chaplygin further applied his theorem to integrate other nonholonomic systems by quadrature \cite{Chap3}, as did Kharlamova later \cite{Ker}. Thus, the reducing multiplier method has historically been interesting and important from the standpoint of the integrability of nonholonomic systems (for more historical notes on the origin of the theorem, see \cite{Sum}, or \cite{E} for a more geometric viewpoint, and \cite{BM2}). This motivation led us to investigate extensions of the theorem in this work in the hope of further expanding its applications to the integrability of nonholonomic systems. One would like a more straightforward procedure which eliminates such guesswork. 

After the introduction of Chaplygin's theorem, subsequent research on the theorem has resulted in, among other things, an extension to the quasicoordinate context \cite{NF}, a study of the geometry behind the theorem \cite{E,Nar1,Nar2}, discoveries of isomorphisms between nonholonomic systems through the use of the theorem \cite{BM}, an example of a system in higher dimensions Hamiltonizable through a similar time reparameterization \cite{FJ3}, an investigation of the necessary conditions for Hamiltonization for abelian Chaplygin systems \cite{Il} (see Section 1.1 for a definition),  Poisson structures for rolling bodies without slipping \cite{BM3,BMK}, and an investigation of rank two Poisson structures in nonholonomic systems \cite{Ram}. In addition, the survey paper \cite{BM2} presents, among other things, many of the known examples to which Chaplygin's theorem is applicable. However, two important aspects yet to be resolved are the extension of the theorem to general nonholonomic systems with symmetry of arbitrary degrees of freedom and, since the theorem rests on the availability of an invariant measure, we are also interested in applying a time reparameterization to ``Hamiltonize'' a nonholonomic system not possessing an invariant measure, where different dynamical effects may arise \cite{BMZ,BM3,BM4}. We note that in this case our resulting ``Hamiltonization'' of a system not possessing an invariant measure should perhaps more properly be called a ``Poissonization,'' since it will in general result in a {\em degenerate} Poisson bracket satisfying the Jacobi identity (see Section 4.5 below), in addition to the continued non-existence of the invariant measure (however, we will continue to refer to this process as ``Hamiltonization,'' keeping in mind this discussion). Moreover, the theorem is commonly used in a rather guess-and-check manner, where one considers systems with known invariant measures and then guesses at the reducing multiplier based on the degrees of freedom. 

In this paper we consider the aforementioned questions for a general nonholonomic system with symmetry governed by the {\em Hamilton-Poincar\'e-d'Alembert equations}. In Section 1 we briefly discuss the mechanics of these systems, as well as for two special cases of them ({\em nonabelian Chaplygin} systems and {\em Euler-Poincar\'e-Suslov} systems), and present results in Sections 2.1, 2.2 and 2.3 which generalize the first part of Chaplygin's Theorem to these higher dimensional systems with symmetry, deriving the necessary conditions (independent of the existence of an invariant measure) for Chaplygin Hamiltonization as a coupled set of first-order partial differential equations in $f$. These equations eliminate the guesswork discussed above, and in the special case of Euler-Poincar\'e-Suslov systems, we present results in Section 2.3 which allow Chaplygin Hamiltonization even when the system does not posses an invariant measure. In Section 3 we use the previous results to extend the idea of conditionally variational systems introduced in \cite{FB} and apply it to Chaplygin Hamiltonize the entire nonholonomic system (reduced constrained equations plus the nonholonomic constraints). Lastly, we devote Section 4 to illustrating these results and showing how special cases of the results presented lead to some of the results found in the works cited above, and discuss some relevant future directions in the Conclusion.


\section{Nonholonomic Systems with Symmetry}


Consider a nonholonomic system with an $n$ dimensional configuration manifold $Q$ and mechanical Lagrangian $L$ which is subject to $k$ linear nonholonomic constraints described by the distribution ${\cal D}$ (moreover, we shall restrict our attention to mechanical Lagrangians for the remainder of the paper). Suppose that we have a Lie group $G$ which acts freely and properly on the configuration space $Q$, with the Lagrangian $L$ and constraints ${\cal D}$ invariant with respect to the induced action of $G$ on $TQ$. For simplicity, assume also that the constraints and the orbit directions span the entire tangent space to the configuration space:

\beg{equation}\label{da}
{\cal D}_{q} + T_{q}\text{Orb}(q)=T_{q}Q, \non
\end{equation}
sometimes known as the \itl{dimension assumption} \cite{B}. 

With sufficient regularity we can use the Legendre transform to pass to the constraint phase space ${\cal M}=\mathbb{F}L({\cal D})$. The quotient space ${\cal \overline{M}}={\cal M}/G$ is a smooth quotient manifold with projection map $\rho: {\cal M} \rightarrow {\cal \overline{M}}$, and all intrinsically defined vector fields then push down to ${\cal M}$, allowing one to write the equations of motion for the reduced constrained Hamiltonian mechanics using a reduced almost-Poisson (in general) bracket on ${\cal \overline{M}}$. The resulting nonholonomic equations of motion are known as the \itl{Hamilton-Poincar\'e-d'Alembert} (HPD) equations and split into a coupled set of second-order equations on the \itl{shape space} $M:=Q/G$ and first-order nonholonomic momentum equations on $\mf{g}^*$ \cite{B,KoMa1997}, whose number equals $s:=dim$ ${\cal S}_{q}$, where ${\cal S}_{q}:={\cal D}_{q} \cap T_{q}\text{Orb}(q)$. 

Following \cite{B} we will now give some of the details, however before doing so let us fix the following index conventions. The indices $a,b,c,\ldots$ will range from $1$ to $k:=dim(\mf{g})$ and correspond to the symmetry directions, $i,j,\ldots$ will range from 1 to $s$ ($s<k$ is the number of momentum equations) and correspond to the symmetry directions along the constraint space, and $\alpha,\beta,\ldots$ will represent the indices for the shape variable $r \in M:=Q/G$ and range from 1 to $m:=n-k=dim(M)$ (the dimension of the shape space). Also, here and for the remainder the of the paper we shall enforce the Einstein summation convention, unless otherwise indicated.

Begin by constructing a body fixed basis $e_{b}(g,r)=Ad_{g}e_{b}(r)$ as in \cite{B}, where $g \in G$ and $r \in M$, such that the infinitesimal generators $(e_{i}(g,r))_{Q}$ of its first $s$ elements at a point $q$ span ${\cal S}_{q}$. Assuming $G$ is a matrix group and $e^{d}_{i}$ is the component of $e_{i}(r)$ with respect to a fixed basis $\{b_{a}\}$ of the Lie algebra $\mf{g}$, we can then represent the constraint distribution ${\cal D}$ as

\beg{equation}\label{HDPD}
{\cal D}=\text{span}\{g^{a}_{d}e^{d}_{i}\partial_{g^{a}},-g^{a}_{b}A^{b}_{\alpha}\partial_{g^{a}}+\partial_{r^{\alpha}}\}, \non
\end{equation}
where we will denote by $\Omega^{i}$ the body angular velocity components of the constrained vertical space. 

Defining the induced coordinates $(g^{a},r^{\alpha},\tilde{p}_{i},\tilde{p}_{\alpha})$ on ${\cal M}$ by

\beg{equation}\label{HDPps}
\tilde{p}_{i} = g^{a}_{d}p_{a}e^{d}_{i}=\mu_{d}e^{d}_{i}, \quad \tilde{p}_{\alpha}=p_{\alpha}-\mu_{b}A^{b}_{\alpha}, \non
\end{equation}

where $\mu \in \mf{g}^*$ and $\mu_{a}$ are its components with respect to a fixed dual basis, $p_{a}=\partial L/\partial \dot{g}^{a}$ and $p_{\alpha}=\partial L/\partial \dot{r}^{\alpha}$, the Hamilton-Poincar\'e-d'Alembert equations on ${\cal \overline{M}}$ are given by \cite{B,BKMM}:

\beg{eqnarray}
\dot{\tilde{p}}_{i} & = & -\mu_{a}C^{a}_{bd}e^{b}_{i}e^{d}_{j}\frac{\partial h_{\cal \overline{M}}}{\partial \tilde{p}_{j}}+\mu_{a}F^{a}_{i\beta}\frac{\partial h_{\cal \overline{M}}}{\partial \tilde{p}_{\beta}}, \label{me} \\
\dot{r}^{\alpha} & = & \frac{\partial h_{\cal \overline{M}}}{\partial \tilde{p}_{\alpha}}, \label{re} \\
\dot{\tilde{p}}_{\alpha} & = & -\frac{\partial h_{\cal \overline{M}}}{\partial r^{\alpha}}-\mu_{a}F^{a}_{j\alpha}\frac{\partial h_{\cal \overline{M}}}{\partial \tilde{p}_{j}}-\mu_{a}{\cal B}^{a}_{\alpha\beta}\frac{\partial h_{\cal \overline{M}}}{\partial \tilde{p}_{\beta}}, \label{pe} 
\end{eqnarray}
along with the constraints 

\beg{equation}\label{hpdce}
\xi^{b} = -{\cal A}^{b}_{\beta}\frac{\partial h_{\cal \overline{M}}}{\partial \tilde{p}_{\beta}}+e^{b}_{j}\frac{\partial h_{\cal \overline{M}}}{\partial \tilde{p}_{j}}.
\end{equation}
Here $h_{\cal \overline{M}}(r,\Omega,\tilde{p})=\tilde{p}_{i}\Omega^{i}+\tilde{p}_{\alpha}\dot{r}^{\alpha}-l_{c}$ is the constrained reduced Hamiltonian (where $l_{c}(r,\dot{r},\Omega)=l(r,\dot{r},\xi=-{\cal A}\dot{r}+\Omega e)$ is the constrained reduced Lagrangian), ${\cal B}^{a}_{\alpha\beta}$ are the coefficients of the curvature of the nonholonomic connection:

\beg{equation}\label{nhcurv}
{\cal B}^{a}_{\alpha\beta} = \frac{\partial {\cal A}^{a}_{\alpha}}{\partial r^{\beta}}-\frac{\partial {\cal A}^{a}_{\beta}}{\partial r^{\alpha}}+C^{a}_{bc}{\cal A}^{b}_{\alpha}{\cal A}^{c}_{\beta}, \non
\end{equation}
where $C^{a}_{bc}$ are the structure constants of the Lie algebra $\mf{g}$, $\xi^{b}=(g^{-1})^{b}_{a}\dot{g}^{a}$, and finally the $F^{a}_{i\beta}$ are given by

\beg{equation}\label{fe}
F^{a}_{i\beta}=\frac{\partial e^{a}_{i}}{\partial r^{\beta}}+C^{a}_{bc}e^{b}_{i}{\cal A}^{c}_{\beta}. \non
\end{equation}

Moreover, in equations (\ref{me})-(\ref{pe}) the quantities $\mu_{a}=\partial l/\partial \xi^{a}$ should be restricted to ${\cal \overline{M}}$ by substituting in the constraints (\ref{hpdce}). Hereafter we shall denote any expression into which the constraints have been substituted with a subscript $c$, as in $(\mu_{a})_{c}$.

Now, the equations (\ref{me})-(\ref{pe}) can be written with respect to an almost-Poisson (AP) bracket $\{\cdot,\cdot\}_{\cal \overline{M}}$ given by \cite{B}

\beg{eqnarray}
\{g,k\}_{\cal \overline{M}} & = & \{\tilde{p}_{i},\tilde{p}_{j}\}\frac{\partial g}{\partial \tilde{p}_{i}}\frac{\partial k}{\partial \tilde{p}_{j}}+\{\tilde{p}_{i},\tilde{p}_{\alpha}\}\left(\frac{\partial g}{\partial \tilde{p}_{i}}\frac{\partial k}{\partial \tilde{p}_{\alpha}}-\frac{\partial g}{\partial \tilde{p}_{\alpha}}\frac{\partial k}{\partial \tilde{p}_{i}}\right) \non \\
& + & \{r^{\alpha},\tilde{p}_{\beta}\}\left(\frac{\partial g}{\partial r^{\alpha}}\frac{\partial k}{\partial \tilde{p}_{\beta}}-\frac{\partial g}{\partial \tilde{p}_{\beta}}\frac{\partial k}{\partial r^{\alpha}}\right)+\{\tilde{p}_{\alpha},\tilde{p}_{\beta}\}\frac{\partial g}{\partial \tilde{p}_{\alpha}}\frac{\partial k}{\partial \tilde{p}_{\beta}}, \label{hpdap}
\end{eqnarray}
where 

\beg{eqnarray}
\{\tilde{p}_{i},\tilde{p}_{j}\} & = & -(\mu_{a})_{c}C^{a}_{bd}e^{b}_{i}e^{d}_{j}, \label{pipj} \\
\{\tilde{p}_{i},\tilde{p}_{\alpha}\} & = & (\mu_{a})_{c}F^{a}_{i\alpha}, \label{pipa} \\
\{r^{\alpha},\tilde{p}_{\beta}\} & = & \delta^{\alpha}_{\beta}, \label{rapb} \non \\
\{\tilde{p}_{\alpha},\tilde{p}_{\beta}\} & = & -(\mu_{a})_{c}{\cal B}^{a}_{\alpha\beta}.\label{papb}
\end{eqnarray}

In Section 2.1 we will derive the necessary and sufficient conditions for the AP bracket (\ref{hpdap}) to become a Poisson bracket after an appropriate choice of quasivelocities (see Section \ref{CHS}). For the rest of Section 2 we will concentrate on achieving that same goal by considering two special cases of the HPD equations (\ref{me})-(\ref{pe}): (1) where ${\cal S}_{q}=\{0\}$, known as the \itl{purely kinematic} or \itl{nonabelian Chaplygin} case,  and (2) the case where $Q=G$, where the resulting equations represent a generalization of the \itl{Euler-Poincar\'e-Suslov} equations \cite{B}. Let us now briefly give the details of these two special cases.


\subsection{Nonholonomic Chaplygin Systems}


Consider the subclass of nonholonomic systems with symmetry corresponding to ${\cal S}_{q}=\{0\}$, known as the {\em purely kinematic case} \cite{B,Koi}, where the group orbits exactly complement the constraints and suppose that $Q \neq G$. These systems are the special case of the HPD equations corresponding to $s=0$ (i.e. dim ${\cal S}_{q}=0$, when there are no nonholonomic momentum equations) and are also known as {\em nonabelian Chaplygin systems} \cite{B,C}. In the special case when $Q=\mathbb{R}^{s} \times S^{r}$ and $G$ is either a torus action $T^{m}$ or acts by translations $\mathbb{R}^{2m}$, they are called {\em abelian Chaplygin systems} and correspond to the classical exposition of Chaplygin systems \cite{NF} where there exist local coordinates $(r^{\alpha},s^{a})$, $\alpha=1,\ldots,n-2m,\;a=n-2m+1,\ldots,n$ such that the Lagrangian $L$ does not depend on the $s^{a}$ coordinates, and where the constraints can be written as $\dot{s}^{a}=-A^{a}_{\alpha}(r)\dot{r}^{\alpha}$. 

Now, from the HPD equations (\ref{me})-(\ref{pe}) we can extract the equations of motion for nonabelian Chaplygin systems as follows. Since $s=0$, we have that $l_{c}=l(r^{\alpha},\dot{r}^{\alpha},-{\cal A}^{a}_{\beta}\dot{r}^{\beta})$, and (assuming sufficient regularity) $h_{\cal \overline{M}}(r,\tilde{p})=\tilde{p}_{\alpha}\dot{r}^{\alpha}-l_{c}$ . Then, equations (\ref{re})-(\ref{pe}) and the constraints (\ref{hpdce}) reduce to:

\beg{eqnarray}
\dot{r}^{\alpha} & = & \frac{\partial h_{\cal \overline{M}}}{\partial \tilde{p}_{\alpha}}, \label{rechap} \\
\dot{\tilde{p}}_{\alpha} & = & -\frac{\partial h_{\cal \overline{M}}}{\partial r^{\alpha}}-(\mu_{a})_{c}{\cal B}^{a}_{\alpha\beta}\frac{\partial h_{\cal \overline{M}}}{\partial \tilde{p}_{\beta}}, \label{pechap}, \\
\xi^{a} & = & -{\cal A}^{a}_{\alpha}(r)\dot{r}^{\alpha}, \label{ce1}
\end{eqnarray}
respectively. For easy reference later on, we also define the semi-basic two-form \cite{CCD} $\Lambda$ on $T^*M$ with components

\beg{equation}\label{lam}
\Lambda_{\alpha\beta}(r,\tilde{p}):=(\mu_{a})_{c}{\cal B}^{a}_{\beta\alpha}, 
\end{equation} 

so that the last term on the right hand side of (\ref{pechap}) can also be expressed as $\Lambda_{\alpha\beta}(\partial h_{\cal \overline{M}}/\partial \tilde{p}_{\beta})$. Moreover, the equations of motion can be written with respect to an AP bracket:

\beg{equation}
\dot{r}^{\alpha} = \{r^{\alpha},h_{\cal \overline{M}}\}_{AP}, \quad \dot{\tilde{p}}_{\alpha} = \{\tilde{p}_{\alpha},h_{\cal \overline{M}}\}_{AP}, \label{apnhe} 
\end{equation}

where the AP bracket is the special case of (\ref{hpdap}) when (\ref{pipj}) and (\ref{pipa}) vanish:

\beg{equation}\label{apbnh}
\{g,k\}^{Chap}_{AP}(r,\tilde{p}) = \{g,k\}_{can}(r,\tilde{p})+\Lambda_{\alpha\beta}\frac{\partial g}{\partial \tilde{p}_{\alpha}}\frac{\partial k}{\partial \tilde{p}_{\beta}},
\end{equation}

for any two functions $g,k:T^*M \rightarrow \mathbb{R}$, where $\{g,k\}_{can}(r,\tilde{p})$ is the canonical Poisson bracket,

\beg{equation}
\displaystyle\{g,k\}_{can}(r,\tilde{p}):=\left(\frac{\partial g}{\partial r_{\alpha}}\frac{\partial k}{\partial \tilde{p}_{\alpha}}-\frac{\partial g}{\partial r_{\alpha}}\frac{\partial k}{\partial \tilde{p}_{\alpha}}\right). \non
\end{equation}


\subsection{Nonholonomic Systems on Lie Groups}


Consider now another special case of the HPD equations corresponding to the setting where the configuration space is the Lie group $G$, so that there is no shape space ($m=dim \; M=0$). The reduced Lagrangian becomes $l=\frac{1}{2}\langle I \xi, \xi \rangle$, where $\xi=g^{-1}\dot{g} \in \mf{g}$ as before, and $I:\mf{g} \mapsto \mf{g}^*$ is the inertia tensor. Substituting in the constraints $\xi^{b}=e^{b}_{j}\Omega^{j}$ we arrive at the reduced constrained Lagrangian $l_{c}(\Omega)$, and assuming sufficient regularity we can define the reduced constrained Hamiltonian $h_{c}(\Omega,\tilde{p})=\tilde{p}_{i}\Omega^{i}-l_{c}$. From (\ref{me}) and (\ref{hpdce}) the equations of motion and constraints then become:

\beg{eqnarray}
\dot{\tilde{p}}_{i} & = & -(\mu_{a})_{c}C^{a}_{bd}e^{b}_{i}e^{d}_{j}\frac{\partial h_{c}}{\partial \tilde{p}_{j}}, \label{EPS} \\
\xi^{b} & = & e^{b}_{j}\frac{\partial h_{c}}{\partial \tilde{p}_{j}}, \label{epsce}
\end{eqnarray}
respectively. These equations are a generalization of the \itl{Euler-Poincar\'e-Suslov} equations \cite{B,KoMa1997}.

We can also write the equations of motion (\ref{EPS}) as:

\beg{equation}\label{eps1}
\dot{\tilde{p}}_{i} = \{\tilde{p}_{i},h_{c}\}^{EPS}_{AP},
\end{equation}
where $\{\cdot,\cdot\}^{EPS}_{AP}$ is the AP bracket:

\beg{equation}\label{apbeps2}
\{g,k\}^{EPS}_{AP} = \{\tilde{p}_{i},\tilde{p}_{j}\}\frac{\partial g}{\partial \tilde{p}_{i}}\frac{\partial k}{\partial \tilde{p}_{j}},
\end{equation}
for any two functions $g,k:(\mf{g}^{c})^* \rightarrow \mathbb{R}$, where $\mf{g}^{c} = \{ \xi \in \mf{g} | \xi^{b}=e^{b}_{j}\Omega^{j}\}$, and where the bracket on the right hand side of (\ref{apbeps2}) is computed by using the canonical bracket on $T^*G$ and then restricting to $\mf{g}^{c}$ (see \cite{B}, Section 5.8 for more details). We note in passing that this bracket is merely (\ref{hpdap}) with (\ref{pipa})-(\ref{papb}) vanishing (since $m=0$). 


\section{Chaplygin Hamiltonization}\label{CHS}


To begin the generalization of Chaplygin's Theorem, we note that one can view Chaplygin's time reparameterization $d\tau=N(q)dt$ from the Introduction in a different way as follows: we have $\dot{q}=dq/dt = N(q)(dq/d\tau)=:N(q)\omega$, which defines the {\em quasivelocities} $\omega$ on $Q$ (For a recent discussion of quasivelocities in nonholonomic mechanics see \cite{BMZ,BM2}). Thus, instead of considering which time reparameterization Hamiltonizes our system, we can rephrase the problem as one of finding a particular set of quasivelocities for which the almost-Poisson bracket (\ref{hpdap}) satisfies the Jacobi identity\footnote{If this is successful, then some authors will also call the nonholonomic system {\em conformally Hamiltonian} (see Section 3.3 in \cite{BM2}).}. To that end, we need to express the AP bracket (\ref{hpdap}) in terms of the quasivelocities $\omega$, to which we now turn. 


\subsection{Chaplygin Hamiltonization of the Hamilton-Poincar\'e-d'Alembert Equations}


Let $j$ be the map $j:(q,\omega_{r},\omega_{\Omega}) \mapsto (q,\dot{r},\Omega)$ and define ${\cal \tilde{P}}=j^*\tilde{p}$. Then locally we have ${\cal L}_{c}(r,\omega)=j^*l_{c}(r,\dot{r},\Omega)=l_{c}(r,\dot{r}^{\alpha}=f\omega^{\alpha},\Omega^{i}=f\omega^{i})$, ${\cal \tilde{P}}_{\alpha}=\partial {\cal L}_{c}/\partial \omega_{r^{\alpha}} = (\partial l_{c}/\partial \dot{r}^{\beta})(\partial \dot{r}^{\beta}/\partial \omega_{r^{\alpha}})$ $=f\tilde{p}_{\alpha}$ and similarly ${\cal \tilde{P}}_{i}=f\tilde{p}_{i}$. Moreover, define the quasivelocities $\omega$ through $\dot{r}=f(r,g)\omega_{r}$ and $\Omega=f(r,g)\omega_{\Omega}$, where $f \in {\cal C}^{1}$ is nonzero on its domain. Then we have the following transformation of the bracket (\ref{hpdap}).

\beg{proposition}\label{6p0}
Consider a nonholonomic system with symmetry governed by the HPD equations (\ref{me})-(\ref{hpdce}). Further, suppose that the matrix $G_{\alpha\beta}:=\partial^{2} l_{c} /\partial \dot{r}^{\alpha} \partial \dot{r}^{\beta}$ is invertible. Then the AP bracket (\ref{hpdap}) becomes the bracket $\{\cdot,\cdot\}'_{\cal \overline{M}}:=(1/f)j^*\{\cdot,\cdot\}_{\cal \overline{M}}$ given by:

\beg{eqnarray}
\{G,K\}'_{\cal \overline{M}} & = & \{{\cal \tilde{P}}_{i},{\cal \tilde{P}}_{j}\}'_{\cal \overline{M}} \frac{\partial G}{\partial {\cal \tilde{P}}_{i}}\frac{\partial K}{\partial {\cal \tilde{P}}_{j}}+\{{\cal \tilde{P}}_{i},{\cal \tilde{P}}_{\alpha}\}'_{\cal \overline{M}} \left(\frac{\partial G}{\partial {\cal \tilde{P}}_{i}}\frac{\partial K}{\partial {\cal \tilde{P}}_{\alpha}}-\frac{\partial G}{\partial {\cal \tilde{P}}_{\alpha}}\frac{\partial K}{\partial {\cal \tilde{P}}_{i}}\right) \non \\
& + & \{r^{\alpha},{\cal \tilde{P}}_{\beta}\}'_{\cal \overline{M}} \left(\frac{\partial G}{\partial r^{\alpha}}\frac{\partial K}{\partial {\cal \tilde{P}}_{\beta}}-\frac{\partial G}{\partial {\cal \tilde{P}}_{\beta}}\frac{\partial K}{\partial r^{\alpha}}\right)+\{{\cal \tilde{P}}_{\alpha},{\cal \tilde{P}}_{\beta}\}'_{\cal \overline{M}} \frac{\partial G}{\partial {\cal \tilde{P}}_{\alpha}}\frac{\partial K}{\partial {\cal \tilde{P}}_{\beta}}, \label{qvapb}
\end{eqnarray}
where 

\beg{eqnarray}
f\{{\cal \tilde{P}}_{i},{\cal \tilde{P}}_{j}\}'_{\cal \overline{M}}  & = & \hat{A}^{k}_{ij}{\cal \tilde{P}}_{k} + \hat{B}^{\gamma}_{ij}{\cal \tilde{P}}_{\gamma}, \label{1} \\
f\{{\cal \tilde{P}}_{i},{\cal \tilde{P}}_{\beta}\}'_{\cal \overline{M}}  & = & \hat{C}^{k}_{i\beta}{\cal \tilde{P}}_{k} + \hat{D}^{\gamma}_{i\beta}{\cal \tilde{P}}_{\gamma}, \label{2} \\
f\{{\cal \tilde{P}}_{\alpha},{\cal \tilde{P}}_{\beta}\}'_{\cal \overline{M}}  & = & \hat{E}^{k}_{\alpha\beta}{\cal \tilde{P}}_{k} + \hat{F}^{\gamma}_{\alpha\beta}{\cal \tilde{P}}_{\gamma}, \label{3} \\
f\{r^{\alpha},{\cal \tilde{P}}_{\beta}\}'_{\cal \overline{M}} & = & f\delta^{\alpha}_{\beta}, \label{4}
\end{eqnarray}
and the components above are given in the Appendix by (\ref{abe})-(\ref{ceps}).
\end{proposition}

\beg{proof} From the reduced Lagrangian $l(r, \dot{r}, \xi)$,

\beg{equation}
l = \frac{1}{2}g_{\alpha\beta}\dot{r}^{\alpha}\dot{r}^{\beta}+g_{a\alpha}\dot{r}^{\alpha}\xi^{a}+\frac{1}{2}g_{ab}\xi^{a}\xi^{b}-V(r), \non
\end{equation} 
we can form the constrained reduced Lagrangian $l_{c}(r,\dot{r},\Omega)$ by substituting in the constraints (\ref{hpdce}) in the form $\xi^{b}=-{\cal A}^{b}_{\beta}\dot{r}^{\beta}+e^{b}_{j}\Omega^{j}$. We then have $\tilde{p}_{\alpha}=\partial l_{c}/\partial \dot{r}^{\alpha}=G_{\alpha\beta}\dot{r}^{\beta}+G^{i}_{\alpha}\tilde{p}_{i}$, where $G_{\alpha\beta}=g_{\alpha\beta}-2g_{a \alpha}{\cal A}^{a}_{\beta}+g_{ab}{\cal A}^{a}_{\alpha}{\cal A}^{b}_{\beta}$, $G^{i}_{\alpha}=\left(g_{b\alpha}-g_{ab}{\cal A}^{a}_{\alpha}\right)\Gamma^{bi}$ and we have used $e^{b}_{j}\Omega^{j}=\Gamma^{bi}\tilde{p}_{i}$, where $\Gamma^{ai}=e^{a}_{j}G^{ji}$, with $G^{ij}$ the inverse of the matrix $G_{ij}=g_{ab}e^{a}_{i}e^{b}_{j}$. Since $G_{\alpha\beta}$ is invertible by assumption (denote its inverse by $G^{\alpha\beta}$), this leads to $\dot{r}^{\gamma}=G^{\gamma\alpha}\tilde{p}_{\alpha}-G^{\gamma\alpha}G^{i}_{\alpha}\tilde{p}_{i}$. Thus, we have:

\beg{eqnarray}
(\mu_{a})_{c} & = & \left(\frac{\partial l}{\partial \xi^{a}}\right)_{c} = M_{a\gamma}\dot{r}^{\gamma}+g_{ab}\Gamma^{bi}\tilde{p}_{i} \non \\
& = & M_{a\gamma}\left(G^{\gamma\alpha}\tilde{p}_{\alpha}-G^{\gamma\alpha}G^{i}_{\alpha}\tilde{p}_{i}\right)+g_{ab}\Gamma^{bi}\tilde{p}_{i} \non \\
& = & M_{a\gamma}G^{\gamma\alpha}\tilde{p}_{\alpha}+\left(g_{ab}\Gamma^{bi}-M_{a\gamma}G^{\gamma\alpha}G^{i}_{\alpha}\right)\tilde{p}_{i}, \non \\
\implies j^*(\mu_{a})_{c} & = & \frac{1}{f}\left[M_{a\gamma}G^{\gamma\alpha}{\cal \tilde{P}}_{\alpha}+\left(g_{ab}\Gamma^{bi}-M_{a\gamma}G^{\gamma\alpha}G^{i}_{\alpha}\right){\cal \tilde{P}}_{i}\right] \label{muqv}
\end{eqnarray}
where we have used the definition of the ${\cal \tilde{P}}$. Then, using the general relation

\beg{equation}
f\{{\cal \tilde{P}}_{I},{\cal \tilde{P}}_{J}\}'_{\cal \overline{M}} = f\{f,\tilde{p}_{J}\}_{\cal \overline{M}}\tilde{p}_{I}-f\{f,\tilde{p}_{I}\}_{\cal \overline{M}}\tilde{p}_{J}+f^{2}\{\tilde{p}_{I},\tilde{p}_{J}\}_{\cal \overline{M}}, \label{brelation}
\end{equation}
which holds for all $I,J=i$ and $I,J=\alpha$, and (\ref{pipj})-(\ref{pipa}) along with (\ref{papb}) and (\ref{muqv}), we get the transformations in (\ref{1})-(\ref{3}). Let us illustrate this for (\ref{1}). 

From (\ref{brelation}) we have:

\beg{eqnarray}
f\{{\cal \tilde{P}}_{i},{\cal \tilde{P}}_{j}\}'_{\cal \overline{M}} & = & f\left[\frac{\partial f}{\partial g^{\sigma}}\frac{\partial \tilde{p}_{j}}{\partial p_{\sigma}}\tilde{p}_{i}-\frac{\partial f}{\partial g^{\sigma}}\frac{\partial \tilde{p}_{i}}{\partial p_{\sigma}}\tilde{p}_{j}\right]+f^{2}\left(-j^*(\mu_{a})_{c}C^{a}_{bd}e^{b}_{i}e^{d}_{j}\right) \non \\
& = & \frac{\partial f}{\partial g^{\sigma}}g^{\sigma}_{d}\left(e^{d}_{j}{\cal \tilde{P}}_{i}-e^{d}_{i}{\cal \tilde{P}}_{j}\right)-f^{2}j^*(\mu_{a})_{c}C^{a}_{bd}e^{b}_{i}e^{d}_{j} \non \\
& = & \overline{C}^{k}_{ij}{\cal \tilde{P}}_{k}+\left(fC^{a}_{bd}e^{b}_{i}e^{d}_{j}M_{a\gamma}G^{\gamma\alpha}G^{k}_{\alpha}-fK^{k}_{ji}\right){\cal \tilde{P}}_{k}+\hat{B}^{\alpha}_{ij}{\cal \tilde{P}}_{\alpha}, \label{demonst}
\end{eqnarray}
where we've used (\ref{muqv}) in the last line of (\ref{demonst}). Indeed, this produces (\ref{1}) and (\ref{abe})-(\ref{bbe}). The remaining transformations (\ref{2})-(\ref{3}) and equations (\ref{cbe})-(\ref{fbe}) follow from similar computations. 

Lastly, we compute $j^*\{r^{\alpha},\tilde{p}_{\beta}\}_{\cal \overline{M}}=f\{r^{\alpha},{\cal \tilde{P}}_{\beta}\}'_{\cal \overline{M}}$ as:

\beg{equation}
f\{r^{\alpha},{\cal \tilde{P}}_{\beta}\}'_{\overline{M}}=\{r^{\alpha},f\tilde{p}_{\beta}\}=\frac{\partial r^{\alpha}}{\partial r^{\gamma}}\frac{\partial (f\tilde{p}_{\beta})}{\partial p_{\gamma}}-\frac{\partial r^{\alpha}}{\partial p_{\gamma}}\frac{\partial (f\tilde{p}_{\beta})}{\partial r^{\gamma}}=f\delta_{\alpha\beta}, \non
\end{equation}
which gives (\ref{4}).
\end{proof}

Proposition \ref{6p0} gives the explicit form for the AP bracket of the HPD equations in terms of the quasivelocities. Now, as stated at the beginning of this section, the idea is to derive the conditions under which the multiplier $f$ makes (\ref{hpdap}) into a Poisson bracket. To that end, we have the first main result.

\beg{theorem}\label{hpdt}
Suppose that we have a nonholonomic system with symmetry satisfying the assumptions of Proposition \ref{6p0} and let $f(r,g)\in {\cal C}^{1}$ be a function which is nonzero everywhere on its domain. Then the almost-Poisson bracket (\ref{hpdap}) is Poisson iff $f$ satisfies:

\beg{eqnarray}
&\hat{B}^{\gamma}_{ij}=0, \quad \hat{D}^{\gamma}_{i \beta}=0, \quad \hat{F}^{\gamma}_{\alpha\beta}=0,& \label{c1} \\
&\hat{A}^{m}_{il}\hat{A}^{l}_{jk}+\hat{A}^{m}_{kl}\hat{A}^{l}_{ij}+\hat{A}^{m}_{jl}\hat{A}^{l}_{ki},& \label{c2} \\
&\hat{C}^{i}_{j\gamma}\hat{E}^{j}_{\alpha\beta}+\hat{C}^{i}_{j\beta}\hat{E}^{j}_{\gamma\alpha}+\hat{C}^{i}_{j\alpha}\hat{E}^{j}_{\beta\gamma}=0,& \label{c3} \\
&\hat{A}^{i}_{kl}\hat{E}^{l}_{\alpha\beta}+\hat{C}^{i}_{l\alpha}\hat{C}^{l}_{k\beta}-\hat{C}^{i}_{l\beta}\hat{C}^{l}_{k\alpha}=0,& \label{c44} \\
&\hat{A}^{l}_{ik}\hat{C}^{k}_{j\alpha}-\hat{C}^{l}_{k\alpha}\hat{A}^{k}_{ij}-\hat{A}^{l}_{jk}\hat{C}^{k}_{i\alpha}=0.& \label{c5} 
\end{eqnarray}
Moreover, the Hamiltonized equations (\ref{me})-(\ref{pe}) become, in this new Poisson bracket $\{\cdot,\cdot\}^{P}_{\cal \overline{M}}$:

\beg{equation}
\dot{r}^{\alpha}=f\{r^{\alpha},{\cal H}_{\cal \overline{M}}\}^{P}_{\cal \overline{M}}, \quad {\cal \dot{\tilde{P}}}_{\alpha}=f\{{\cal \tilde{P}_{\alpha}},{\cal H}_{\cal \overline{M}}\}^{P}_{\cal \overline{M}}, \quad {\cal \dot{\tilde{P}}}_{i}=f\{{\cal \tilde{P}}_{i},{\cal H}_{\cal \overline{M}}\}^{P}_{\cal \overline{M}}, \label{hamhpd} 
\end{equation}
where ${\cal H}_{\cal \overline{M}}=j^*h_{\cal \overline{M}}$. 
\end{theorem}

\beg{proof}
In order for (\ref{qvapb}) to become a Poisson bracket it must satisfy the Jacobi identity. We can compute the results of this restriction, which then leads to the conditions (\ref{c1})-(\ref{c5}) that $f$ must satisfy. For better readability of this paper, we have left these computations to the second section of the Appendix. As for the second half of the Theorem, we simply note that since $j^*r=r$, $j^*\tilde{p}={\cal \tilde{P}}$ and $j^*h_{\cal \overline{M}}={\cal H}_{\cal \overline{M}}$, then the AP bracket representation of the equations of motion (\ref{me})-(\ref{pe}) becomes $j^*(r^{\alpha}-\{r^{\alpha},h_{\cal \overline{M}}\}_{\cal \overline{M}})=0$, which gives the first equation in (\ref{hamhpd}) since $j^*\{\cdot,\cdot\}_{\cal \overline{M}}=f\{\cdot,\cdot\}^{P}_{\cal \overline{M}}$, and similarly for the remaining. 
\end{proof}

Theorem \ref{hpdt} represents the necessary conditions under which a given nonholonomic system with symmetry admitting a representation within the HPD framework can be Chaplygin Hamiltonized into (\ref{hamhpd}). It is a generalization of Chaplygin's Theorem not only to higher dimensional nonholonomic systems with symmetry, but also in that it does not presuppose the existence of an invariant measure. Indeed, although the conditions (\ref{c1})-(\ref{c5}) seem rather involved they are no more than a coupled set of first-order partial differential equations in $f$ which can be solved using any of the popular mathematical software packages. Thus, contrary to the traditional usage of Chaplygin's Theorem found in the literature (where the reducing multiplier is typically guessed at by knowing the system's invariant measure and degrees of freedom), Theorem \ref{hpdt} does not require the invariant measure and eliminates the guesswork. In fact, as we will show below, for nonabelian Chaplygin systems we will recover the second part of Chaplygin's Theorem from the analogous $f$ conditions in that case, showing that for these types of nonholonomic systems it is more advantageous to solve the corresponding $f$ conditions instead of guessing at the multiplier using the invariant measure density, since if a solution exists then we will get the invariant measure density for free. Moreover, as we will also show below and have already mentioned above, one can now also apply Chaplygin Hamiltonization to systems for which both Chaplygin's Theorem and the guesswork above are inapplicable: nonholonomic systems which do not possess an invariant measure density. The Chaplygin sleigh (which we discuss in Section \ref{CSS}) is perhaps the best illustration. 

Let us now turn to the Chaplygin Hamiltonization of the two special cases considered in Sections 1.1 and 1.2 above. Indeed, consider now the special cases of the HPD equations given by the nonabelian Chaplygin nonholonomic systems, as in Section 1.1, and the Euler-Poincar\'e-Suslov case of Section 1.2, where there is no shape space. From Theorem \ref{6p0}, in these cases the quasivelocity transformations read $\dot{r}^{\alpha}=f(r)\omega_{r^{\alpha}}$ (we shall write this simply as $\dot{r}=f(r)\omega$ henceforth) and $\Omega^{i}=f(g)\omega_{\Omega^{i}}$ (we shall write this simply as $\Omega=f(g)\omega$ henceforth), respectively, where $f \in {\cal C}^{1}$ is nonzero on its domain. Moreover, defining the maps $j_{r}:(q,\omega) \mapsto (q,\dot{r})$ and $j_{\Omega}:(\omega) \mapsto (\Omega)$ we have the following Corollary of Theorem \ref{hpdt}.

\beg{corollary}\label{hpdc}
(1) For a nonabelian Chaplygin nonholonomic system $(L,G,{\cal D})$ described by (\ref{rechap})-(\ref{ce1}) with almost-Poisson formulation (\ref{apnhe})-(\ref{apbnh}) satisfying the assumptions of Proposition \ref{6p0}, the necessary and sufficient conditions for Chaplygin Hamiltonization (using $d\tau = f dt$) on $j_{r}^*M$ are that $f$ satisfy  

\beg{equation}\label{geocondc}
j_{r}^*\{g,k\}^{Chap}_{AP}(r,p) = f\{G,K\}_{can}(r,{\cal \tilde{P}}),
\end{equation}

for all $\alpha,\nu,\delta= 1,\ldots,m$ (recall from Section 1 that $m=n-k=dim(M)$), and where $\{\cdot,\cdot\}_{can}$ is the canonical bracket on $j_{r}^*M$. 

(2) For an Euler-Poincar\'e-Suslov nonholonomic system described by (\ref{EPS})-(\ref{epsce}) with  almost-Poisson formulation (\ref{eps1})-(\ref{apbeps2}) satisfying the assumptions of Proposition \ref{6p0}, the necessary and sufficient conditions for Chaplygin Hamiltonization (using $d\tau = f dt$) on $j^*_{\Omega}(\mf{g}^{c})^*$ are that $f$ satisfy 

\beg{equation}\label{geoconde}
j^*_{\Omega}\{g,k\}^{EPS}_{AP}=f\{G,K\}_{-}^{j^*_{\Omega}(\mf{g}^{c})^*},
\end{equation}

where $\{\cdot,\cdot\}_{-}^{j^*_{\Omega}(\mf{g}^{c})^*}$ is the (minus) Lie-Poisson bracket on $j^*_{\Omega}(\mf{g}^{c})^*$. 

Equivalently, in local form conditions (\ref{geocondc}) and (\ref{geoconde}) read

\beg{eqnarray}
&\displaystyle \frac{\partial f}{\partial r^{\delta}}G_{\alpha\nu}+\frac{\partial f}{\partial r^{\nu}}G_{\alpha\delta}-2\frac{\partial f}{\partial r^{\alpha}}G_{\delta\nu}=f\left(K^{\mu}_{\alpha\delta}G_{\mu\nu}+K^{\mu}_{\alpha\nu}G_{\mu\delta}\right),&  \label{condhdf} \\
&S^{l}_{km}S^{m}_{ij}+S^{l}_{jm}S^{m}_{ki}+S^{l}_{im}S^{m}_{jk}=0, \quad \forall i,j,k,l=1,\ldots,n-k,& \label{epslocal} 
\end{eqnarray}

respectively, where $S^{l}_{km}:=-\left(K^{l}_{mk}-\overline{C}^{l}_{km}\right).$
\end{corollary}

\beg{proof} (1) Let us consider the nonabelian Chaplygin case first. Since this case corresponds to the situation in which $s=dim \; {\cal S}_{q}=0$, then only Greek indices survive in Theorem \ref{hpdt}. Moreover, using that in this case $g_{ab}=g_{a\alpha}=0$ for all $a,\alpha$ and $f$ is independent of $g$, we can extract the relevant Hamiltonization conditions from (\ref{c1})-(\ref{c5}). The only non-vacuous condition amongst those in Theorem \ref{hpdt} is then $\hat{F}^{\gamma}_{\alpha\beta}=0$. From (\ref{fbe}) this leads to the condition (\ref{geocondc}), and its local form in (\ref{condhdf}).

(2) For the Euler-Poincar\'e-Suslov case, since this corresponds to the special case of the HPD equations in which $m=dim \; M=0$, then only the Latin indices survive in Theorem \ref{hpdt}. Thus, $g_{\alpha\beta}=g_{a\alpha}=0$ for all $a,\alpha,\beta$, and since $f$ is independent of $r$, the only non-vacuous condition amongst those in Theorem \ref{hpdt} is condition (\ref{c2}). However, note that because of the fact that $m=0$, the first term in (\ref{abe}) vanishes, giving the condition (\ref{geoconde}), and its local form in (\ref{epslocal}).
\end{proof}

For completeness, we should note that based on the results of Corollary \ref{hpdc} we can write the reduced constrained mechanics of a nonabelian Chaplygin and Euler-Poincar\'e-Suslov nonholonomic system from (\ref{hamhpd}) as

\beg{eqnarray}
& \dot{r}^{\alpha} = f\{r^{\alpha},{\cal H}_{\cal \overline{M}}\}_{can} \quad \text{and} \quad {\cal \dot{\tilde{P}}}_{\beta} = f\{{\cal \tilde{P}}_{\beta},{\cal H}_{\cal \overline{M}}\}_{can}, & \label{qve} \\
& {\cal \dot{\tilde{P}}}_{i}=f\{{\cal \tilde{P}}_{i},{\cal H}_{\cal \overline{M}}\}_{-}^{j^*_{\omega}(\mf{g}^{c})^*},& \label{epsqve}
\end{eqnarray}
respectively. The reader familiar with the usual treatment of Chaplygin's theorem will note the absence of the reparameterized time $\tau$ in (\ref{qve}). In fact, in the context of Chaplygin's work, as well as to compare directly with \cite{BM}, we note that the quasi-Hamiltonian forms (\ref{qve})-(\ref{epsqve}) are the ``$t$-time'' analogues of the Hamiltonian forms stated in the classical Chaplygin Reducibility Theorem in ``$\tau$-time,'' and the two are related through $\dot{r}=f r'$, ${\cal \dot{\tilde{P}}}=f{\cal \tilde{P}}'$, where $r'=dr/d\tau$ and ${\cal \tilde{P}}'=d{\cal \tilde{P}}/d\tau$.

Now, given the more general conditions in Corollary \ref{hpdc} (which are valid for nonabelian Chaplygin and Euler-Poincar\'e-Suslov systems in arbitrary degrees of freedom), let us proceed to extract both parts of Chaplygin's Reducing Multiplier Theorem as special cases.


\subsection{Chaplygin's Reducing Multiplier Theorem}


We now specialize to the case when $m=dim \; M=2$ (the two degree of freedom case) to extract the first part of Chaplygin's Theorem.

\beg{corollary}\label{c4}
The necessary and sufficient condition for a Chaplygin nonholonomic system $(L,G, {\cal D})$ in two degrees of freedom ($m=2$) to be Chaplygin Hamiltonizable is that

\beg{equation}
\frac{\partial K^{1}_{12}}{\partial r^{1}} = - \frac{\partial K^{2}_{12}}{\partial r^{2}}, \label{2dofca}
\end{equation}

or equivalently that the system (\ref{rechap})-(\ref{pechap}) possess a nonzero invariant measure density $N(r) \in {\cal C}^{1}$. The multiplier is then given by $f(r) = e^{\int K^{1}_{12} dr^{2}}=N$.\end{corollary}

\beg{proof} From Corollary \ref{hpdc}, the only independent conditions in (\ref{geocondc}) in the two degree of freedom case are:

\beg{eqnarray}
&\displaystyle\left(\frac{\partial f}{\partial r^{2}}-fK^{1}_{12}\right)G_{11}-\left(\frac{\partial f}{\partial r^{1}}+fK^{2}_{12}\right)G_{12}=0,& \label{2df2} \non \\
&\displaystyle\left(\frac{\partial f}{\partial r^{2}}-fK^{1}_{12}\right)G_{21}-\left(\frac{\partial f}{\partial r^{1}}+fK^{2}_{12}\right)G_{22}=0.& \non \label{2df3}
\end{eqnarray}

Since we have assumed that $G_{\alpha\beta}$ is invertible, the necessary and sufficient condition for the satisfaction of these equations is that the parenthetical terms vanish. The resulting set of equations is soluble iff (\ref{2dofca}) is satisfied, in which case $f$ is given in explicit form as in the Corollary. However, as we showed in \cite{FB}, if the the constrained reduced system (\ref{rechap})-(\ref{pechap}) has an invariant measure, then its density $N$ (for $m=2$) satisfies:

\beg{equation}\label{imde}
K^{1}_{12}=\frac{1}{N}\frac{\partial N}{\partial r^{2}}, \quad K^{2}_{12}=-\frac{1}{N}\frac{\partial N}{\partial r^{1}}. \non
\end{equation}

One sees immediately that this satisfies (\ref{2dofca}), and hence $f=N$ is a multiplier. \end{proof}

Moving on to the second part of Chaplygin's Theorem, the Proposition below shows that it too follows from the Hamiltonization condition (\ref{geocondc}).

\begin{proposition}\label{6p4}
Suppose $f$ satisfies the conditions of Corollary \ref{hpdc}. Then the original system (\ref{rechap})-(\ref{pechap}) has an invariant measure with density $f^{m-1}$. 
\end{proposition}

\beg{proof} Suppose $f$ satisfies (\ref{condhdf}). Multiplying by $\dot{r}^{\delta}$ and $\dot{r}^{\nu}$ and adding results in:

\beg{equation}
\frac{1}{f}\left(\frac{\partial f}{\partial r^{\beta}}p_{\alpha}-\frac{\partial f}{\partial r^{\alpha}}p_{\beta}\right)\dot{r}^{\beta}=\left(\frac{\partial l}{\partial \xi^{a}}\right)_{c}{\cal B}^{a}_{\alpha\beta}\dot{r}^{\beta}. \non
\end{equation}

Comparing the $\dot{r}^{\beta}$ coefficients yields:

\beg{equation}\label{a1a}
\Lambda_{\beta\alpha}=\frac{1}{f}\left(\frac{\partial f}{\partial r^{\beta}}p_{\alpha}-\frac{\partial f}{\partial r^{\alpha}}p_{\beta}\right),
\end{equation}

where we remind the reader of the definition of $\Lambda$ from (\ref{lam}), and have used $p_{\alpha}=\partial l_{c} / \partial \dot{r}^{\alpha}$ (we will drop the tildes in $p$ here). We also note that the relationship (\ref{a1a}) was also presented as a sufficient condition for the existence of an invariant measure by \cite{Stan} (see also \cite{C}), but here is derived from the conditions of Corollary \ref{hpdc}. Thus, for a Chaplygin Hamiltonizable system the second term on the right hand side of (\ref{pechap}) can be written in terms of $f$ as in (\ref{a1a}). 

Now, suppose $X_{nh}=\dot{r}^{\alpha}\partial_{r^{\alpha}}+\dot{p}_{\alpha}\partial_{p^{\alpha}}$ is the nonholonomic vector field solution to the system (\ref{rechap})-(\ref{ce1}). We will show that $f^{m-1}$ is an invariant measure density by showing that the vector field $f^{m-1}X_{nh}$ has zero divergence. A straightforward calculation yields

\beg{equation}\label{a2a}
\text{div}\;(f^{m-1}X_{nh}) = \frac{\partial (f^{m-1}\dot{r}^{\alpha})}{\partial r^{\alpha}}+\frac{\partial (f^{m-1}\dot{p}_{\alpha})}{\partial p_{\alpha}}=f^{m-2}\dot{r}^{\alpha}\left((m-1)\frac{\partial f}{\partial r^{\alpha}}+f\frac{\partial \Lambda_{\beta\alpha}}{\partial p^{\beta}}\right),
\end{equation}

and a simple calculation of the last term in (\ref{a2a}) using (\ref{a1a}) then shows that the divergence does indeed vanish and completes the proof. \end{proof}

Corollary \ref{hpdc} yields the necessary conditions for the Chaplygin Hamiltonization of the nonabelian Chaplygin system (\ref{rechap})-(\ref{ce1}), which locally are the first-order partial differential equations (\ref{condhdf}) in $r$. Proposition \ref{6p4} then completes the generalization by providing us with the invariant measure density given a solution to (\ref{geocondc}). We should stress, however, that the converse of Proposition \ref{6p4} is not true in general. That is, given an $m>2$ degree of freedom nonholonomic system with an invariant measure, its Chaplygin multiplier $f$ may or may not coincide with the invariant measure density (or any other smooth function of it). This is most easily seen by first assuming that the $m$ degree of freedom Chaplygin system has an invariant measure with density $f^{m-1}$, so that the right hand side of (\ref{a2a}) vanishes. Inserting the resulting equation for $\partial f / \partial r^{\alpha}$ into (\ref{condhdf}) yields

\beg{equation}\label{co1}
2G_{\delta\nu}K^{\beta}_{\alpha\beta}-\left(G_{\alpha\nu}K^{\beta}_{\delta\beta}+G_{\alpha\delta}K^{\beta}_{\nu\beta}\right)=(m-1)\left(K^{\mu}_{\alpha\delta}G_{\mu\nu}+K^{\mu}_{\alpha\nu}G_{\mu\delta}\right).
\end{equation}

The conditions in (\ref{co1}) are, for $m>2$, conditions solely arising from the nonholonomic system itself, as (\ref{co1}) depends only on the metric of the Lagrangian and the curvature of the connection (for $m=2$ (\ref{co1}) is vacuous, a manifestation of Corollary \ref{c4})\footnote{Moreover, it should also be clear that for $m>2$, using any smooth function $F(f;m)$ will also lead to conditions similar to (\ref{co1}).}. Thus, if the metric and curvature of the connection of a Chaplygin nonholonomic system interact precisely as in (\ref{co1}) then the converse of Proposition \ref{6p4} holds. However, given the rarity of such an event, we believe that the ordering of the Hamiltonization process for nonabelian Chaplygin systems that first begins by attempting to solve the conditions (\ref{condhdf}) and then extracting the invariant measure density from Proposition \ref{6p4} is best\footnote{Nonetheless, it is impressive that some authors \cite{FJ,FJ2,FJ3} have effectively found nonholonomic systems for which (\ref{co1}) is satisfied.}.

In another direction, it may be the case, however, that (\ref{condhdf}) does not have a solution. This does not mean that the system is not Chaplygin Hamiltonizable though, since it may still possess more symmetries which further reduce the degrees of freedom, and which allow one to seek such a solution on the second reduced phase space. We illustrate such a situation in the next section, making use of the classical Routhian \cite{B} to explore the effect of additional simple symmetries in (\ref{rechap})-(\ref{pechap}) on its Chaplygin Hamiltonizability.

\subsection{Momentum Conservation and Chaplygin Hamiltonization}\label{MCS}


Suppose that (\ref{condhdf}) has no solutions, but that the nonholonomic system possesses momentum conservation laws that we have yet to account for. With the aid of these conservation laws, we can apply the reduction process to further reduce the degrees of freedom of the system and re-attempt a Hamiltonization on the second reduced space. 

In order to illustrate this in a simple manner, we restrict ourselves in this section to Chaplygin systems which we will call {\em nonholonomic cylic}. By this we mean that we have an abelian Lie group $H$ acting on $M=Q/G$ by $M \ni r^{\alpha}=(w^{\alpha'},v^{i}) \mapsto (w^{\alpha'},v^{i}+h^{i})$, $h \in H$, where $i=1,\ldots l=dim (H)$ and $\alpha'=1,\ldots,m-l$ and such that $\Lambda_{\alpha'i}=0 \; \forall i,\alpha'$ (we shall hereafter denote the nonconserved conjugate variable indices $w$ with a prime) and such that the action leaves the Lagrangian and constraints invariant. Under these assumptions the $v^{i}$ equations in (\ref{pechap}) lead to the momentum conservation laws\footnote{Since the $H$-invariance implies that $l_{c}$ does not depend explicitly on the $v^{i}$, these variables are {\em cyclic} and produce momentum conservation laws in unconstrained systems. However, due to the presence of the $\Lambda_{\alpha\beta}$, cyclic variables are not enough to produce the conservation laws, hence the introduction of the terminology ``nonholonomic cyclic.''} and we can thus set $p_{i}=\lambda_{i}=\;$constant and perform a partial Legendre transform in the $v^{i}$ variables to form, analogous to the classical Routhian \cite{MR,CMR2}, the constrained Routhian $R_{c}(w,\dot{w})$ defined by

\beg{equation}
R_{c}(w,\dot{w}) :=\left[l_{c}(w,\dot{w},\dot{v})-\lambda_{i}\dot{v}^{i}\right]_{p_{i}=\lambda_{i}}. \label{routh}
\end{equation}

Now, using the well-known fact \cite{MR} that the Euler-Lagrange expressions of the nonconserved variables of $l_{c}$ are equivalent to the Euler-Lagrange expressions of the nonconserved variables of $R_{c}$, we can write the Lagrange-d'Alembert equations for $R_{c}$ as 

\beg{equation}
\frac{d}{dt}\frac{\partial {R}_{c}}{\partial \dot{w}^{\alpha'}}-\frac{\partial {R}_{c}}{\partial w^{\alpha'}}=-\left(\frac{\partial l}{\partial \xi^{a}} \right)_{c}{\cal B}^{a}_{\alpha'\beta'}\dot{w}^{\beta'}, \label{re1}
\end{equation}
along with the conservation equations $\dot{p}_{i}=0$. Furthermore, the last term on the right hand side of (\ref{re1}) can be rewritten in terms of the Routhian:

\beg{eqnarray}
\left(\frac{\partial l}{\partial \xi^{a}} \right)_{c}{\cal B}^{a}_{\alpha'\beta'} &=& M_{a\alpha}G^{\alpha\beta}\frac{\partial l_{c}}{\partial \dot{r}^{\beta}}{\cal B}^{a}_{\alpha'\beta'}= \left[M_{a\alpha}G^{\alpha\epsilon'}\frac{\partial l_{c}}{\partial \dot{w}^{\epsilon'}}+M_{a\alpha}G^{\alpha i}\lambda_{i}\right]{\cal B}^{a}_{\alpha'\beta'}, \non \\
&=& \left[\left(M_{a\alpha}G^{\alpha\epsilon'}\frac{\partial  R_{c}}{\partial \dot{w}^{\epsilon'}}\right)+M_{a\alpha}G'^{\alpha i}\lambda_{i}\right]{\cal B}^{a}_{\alpha'\beta'}, \label{a1} \\
&=& K^{\epsilon'}_{\alpha'\beta'}\frac{\partial R_{c}}{\partial \dot{w}^{\epsilon'}}+K^{i}_{\alpha'\beta'}\lambda_{i}, \label{a2}
\end{eqnarray}

where $G'^{\alpha i}=G^{\alpha i}-G^{\alpha\epsilon'}\overline{G}^{ij}G_{j\epsilon'}$, assuming the invertibility of $\overline{G}^{ij}:=(\partial^{2} l_{c}/\partial v^{i}\partial v^{j})$ as well as that of the kinetic energy matrix of $l_{c}$ (we also remind the reader of the definition of the $K^{\epsilon}_{\alpha\beta}$ in (\ref{kchap})). Moreover, we shall henceforth denote the parenthetical term in (\ref{a1}) by $(\mathbb{F}R_{c})'$.

Then, since $R_{c}$ can now be interpreted as a function on $T(M/H)$, we can now attempt to Hamiltonize (\ref{re1}) on this second reduced space. To that end, defining the maps $\overline{j}:(w,\omega) \mapsto (w,\dot{w})$ and $\overline{j}_{p}:(r,\lambda_{i}) \mapsto (r,p_{i})$, we have the second main result.

\beg{theorem}\label{6t6}
Suppose that the nonabelian Chaplygin nonholonomic system given by (\ref{rechap})-(\ref{ce1}) is not Hamiltonizable by Corollary \ref{hpdc} but is nonholonomic cyclic. Further, suppose that the kinetic energy matrix of $l_{c}$ and the sub-matrix $\overline{G}^{ij}:=(\partial^{2} l_{c}/\partial v^{i}\partial v^{j})$ are invertible. Then if there exists a multiplier $f(w)$, nonzero everywhere on its domain with $f(w) \in {\cal C}^{1}$, satisfying 

\beg{equation}
\left[\omega,\frac{\partial {\cal R}_{c}}{\partial \omega}\right]^{\ast}=\langle (\mathbb{F}{\cal R}_{c})',{\cal B}(\omega,f\partial_{w})\rangle,\label{rthm}
\end{equation}

where ${\cal R}_{c}(w,\omega)=\overline{j}^*R_{c}$ and $(\mathbb{F}{\cal R}_{c})'=\overline{j}^*(\mathbb{F}R_{c})'$, the reduced system is Chaplygin Hamiltonizable on $M^\prime=M/H$ under the choice of quasivelocity $\dot{w}=f\omega$. Furthermore, assuming sufficient regularity its dynamics on $T(M/H)$ can be written in the quasi-Hamiltonian form

\beg{equation}
\dot{w}^{\alpha'} = f\{w^{\alpha'},{\cal H}_{M'}\}_{AP}, \quad {\cal \dot{\tilde{P}'}}_{\beta'} = f\{{\cal \tilde{P}'}_{\beta'},{\cal H}_{M'}\}_{AP}, \label{rre}
\end{equation}

where ${\cal \tilde{P}'}_{\alpha'}=\partial {\cal R}_{c} / \partial \omega^{\alpha'}$ and ${\cal H}_{M'}=\omega^{\alpha'}{\cal \tilde{P}}'_{\alpha'}-{\cal R}'_{c}|_{\omega \rightarrow {\cal \tilde{P}'}}$ is the Hamiltonian and the almost-Poisson bracket is defined by:

\beg{equation}
\{G,K\}_{AP}(w',{\cal \tilde{P}}') = \{G,K\}_{can}(w',{\cal \tilde{P}}')-fK^{i}_{\alpha'\beta'}\lambda_{i}\frac{\partial G}{\partial {\cal \tilde{P}}'_{\alpha'}}\frac{\partial K}{\partial {\cal \tilde{P}}'_{\beta'}}, \label{PB2}
\end{equation}
where $\{\cdot,\cdot\}_{can}$ is the canonical bracket on $T(M/H)$. Moreover, the bracket automatically satisfies the Jacobi identity for $dim(M')=2$.
\end{theorem}

\beg{proof} Under the quasivelocity transformation $\dot{w}=f(w)\omega$ the constrained reduced equations (\ref{re1}) become (taking into account (\ref{a2})):

\beg{equation}
\frac{d}{dt}\frac{\partial {\cal R}_{c}}{\partial \omega^{\alpha'}}-f\frac{\partial {\cal R}_{c}}{\partial w^{\alpha'}}=\left(W^{\epsilon'}_{\alpha'\beta'}\frac{\partial {\cal R}_{c}}{\partial \omega^{\epsilon'}}\omega^{\beta'}\right)-f^{2}K^{i}_{\alpha'\beta'}\lambda_{i}\omega^{\beta'}, \label{a23}
\end{equation}

where $W^{\epsilon'}_{\alpha'\beta'}:=fK^{\epsilon'}_{\beta'\alpha'}-C^{\epsilon'}_{\alpha'\beta'}$. Now, if $f$ is chosen to satisfy (\ref{rthm}), then the parenthetical term in (\ref{a23}) vanishes. By defining the Hamiltonian ${\cal H}_{M'}$ as in the statement of the Theorem, the equations of motion can then be written as in (\ref{rre}) with the almost-Poisson bracket (\ref{PB2}). Lastly, a straightforward computation shows that the Jacobi identity is automatically satisfied for dim$(M')=2$, owing to the fact that the non-canonical part in (\ref{PB2}), $\{{\cal P}_{\alpha'},{\cal P}_{\beta'}\}_{AP}$, is independent of the momenta. \end{proof}

Theorem \ref{6t6} will be used below when discussing the Chaplygin sphere (a classic example of how the apparent failure of Hamiltonizability can be reversed in the presence of momentum conservation laws) and the Snakeboard as well. However, let us remark here that in \cite{BM} the authors consider an extension of the Chaplygin method to the case where gyroscopic forces are involved, analyzing the Hamiltonization of nonholonomic systems in two degrees of freedom. In their subsequent bracket description of the Hamiltonized mechanics there appear non-canonical parts which in this paper manifest themselves as the second term in (\ref{PB2}), resulting from the second reduction to $M/H$. Indeed, we can compare the results of Theorem \ref{6t6} to the exposition in \cite{BM} by first noting that the non-canonical part of the bracket (\ref{PB2}) is the many degree of freedom analogue of $\overline{S}$ in equation (3) of \cite{BM}. This is best seen by defining $\overline{S}_{\alpha' \beta'}:=-fK^{i}_{\alpha' \beta'}\lambda_{i}$ along with the 2-form $\Omega:=\overline{S}_{\alpha' \beta'}dw^{\alpha'} \wedge dw^{\beta'}$. The two-form $\Omega$ is exact when $dim \; M =2$, and in that case (or any other case when it is exact), one can then locally write $\Omega=d\beta$, where $\beta = W_{\alpha'}(w)dw^{\alpha'}$, and as the authors in \cite{BM} point out, the constrained reduced equations (\ref{rre}) can then be rewritten as 

\beg{equation}
\frac{d}{dt}\frac{\partial {\cal R}_{W}}{\partial \omega^{\alpha'}}-f\frac{\partial {\cal R}_{W}}{\partial w^{\alpha'}}=0, \label{a24}
\end{equation}

where $R_{W}(w,\omega)=R_{c}(w,\omega)+W_{\alpha'}(w)\omega^{\alpha'}.$ 

As a preliminary application of the results above, we shall now use Corollary \ref{hpdc} to extend prior Hamiltonization results from \cite{FB} of a class of nonholonomic systems known as conditionally variational systems, which are nonholonomic systems that can be Hamiltonized in full subject to the imposition of initial conditions that satisfy the constraints.

\section{Conditionally Variational Systems in the Quasivelocity Context}


In \cite{FB} we discussed the notion of a \textit{conditionally variational} nonholonomic system. Briefly, these systems have the property that the constrained Euler-Lagrange equations are \textit{Lagrangian}, hence making it possible to express the constrained dynamics in a variational manner. However, we showed that under certain additional requirements for the original system's Lagrangian one can construct the ``variational'' Lagrangian $L_{V}$ whose Euler-Lagrange equations reproduce the nonholonomic equations when the initial conditions are chosen to satisfy the constraints (as they must anyway). Hence, such nonholonomic systems can be realized as \textit{variational} systems provided the initial conditions satisfy the constraints and that the original Lagrangian satisfies certain requirements. 

In that paper we showed that such a realization was possible only in the cases when the nonholonomic system possessed an invariant measure with constant density $N(r)$ (the vertical rolling disk of Section \ref{VRDS} is such a system). However, using Corollary \ref{hpdc} we can now extend the results in \cite{FB} to a more general setting if we instead focus on the Chaplygin Hamiltonized system. To that end we have the following result:

\beg{theorem}\label{t8}
Suppose that for a given abelian Chaplygin nonholonomic system $(L,G,{\cal D})$ with constraints given by

\beg{equation}\label{spare}
\phi^{a}(q,\dot{q})=\dot{s}^{a}+A^{a}_{\alpha}(r)\dot{r}^{\alpha}, \quad a=1,\ldots,k<n, \non
\end{equation}

where $q=(r,s)$, we have found an $f$ as in Corollary \ref{hpdc} above and let ${\cal L}(q,\omega):=L(q,\dot{r}=f\omega,\dot{s}=f\omega)$ and $\phi^{a}(q,\omega)=\phi^{a}(q,\dot{r}=f\omega,\dot{s}=f\omega)$. Then if the matrix $\widetilde{g}_{ab}:=(\partial^{2} {\cal L}/\partial \omega^{a}\partial \omega^{b})$ is invertible, the nonholonomic mechanics of the original system can be derived from the (almost) Euler-Lagrange equations

\beg{equation}
\frac{d}{dt}\frac{\partial {\cal L}_{V}}{\partial \omega^{I}}-f\frac{\partial {\cal L}_{V}}{\partial q^{I}} = 0, \quad\quad I=1,\ldots,n, \label{qH} \\
\end{equation}

by using the Lagrangian ${\cal L}_{V}(q,\omega)$ defined by 

\beg{equation}
{\cal L}_{V}(q,\omega) = {\cal L}(q,\omega) - \frac{1}{f}\frac{\partial {\cal L}}{\partial \omega^{a}}\phi^{a}(q,\omega), \label{lv}
\end{equation}

and imposing the nonholonomic constraints initially.

\end{theorem}

\beg{proof} The existence of an $f$ which Hamiltonizes the system guarantees, by part (2) of Proposition 3 in \cite{FB}, that the system $({\cal L}(q,\omega),\phi(q,\omega))$ is conditionally variational after the reparameterization $d\tau = f(r)dt$. Then, the Theorem follows by Proposition 5 of \cite{FB} again. \end{proof}

Theorem \ref{t8} extends Chaplygin's Theorem in a different direction. Unlike Corollary \ref{hpdc}, it gives one a method to Hamiltonize the \textit{entire} system (similar to some of our earlier work \cite{BFM}) after a time reparameterization. Although it is impossible for a nonholonomic system to be Hamiltonian \cite{B}, Theorem \ref{t8} begins to answer the open problem briefly discussed in \cite{BM2} and elsewhere of lifting the Hamiltonization of the reduced problem to the whole system. We illustrate this and our other results below by applying these ideas to some common and well-known nonholonomic systems.


\section{Examples}


The simplest illustrations of the above results can be found in low dimensions, specifically the two degree of freedom case. Although this is the original setting for Chaplygin's Reducibility Theorem, we will discuss the Chaplygin sleigh (which, due to its lack of an invariant measure, cannot be handled by Chaplygin's Theorem), among other things, and also take this opportunity to illustrate Theorem \ref{t8} as well. 


\subsection{The Vertical Rolling Disk}\label{VRDS}


Consider the nonholonomic vertical rolling disk pictured in Figure 1 below with configuration space $Q=\mathbb{R}^{2}\times S^{1} \times S^{1}$ and parameterized by the coordinates $(x,y,\theta,\varphi)$, where $(x,y)$ is the position of the center of mass of the disk, $\theta$ is the angle that a point fixed on the disk makes with respect to the vertical, and $\varphi$ is measured from the positive $x$-axis. This system has Lagrangian and constraints given by:

\beg{eqnarray}
L & = & \frac{1}{2}m(\dot{x}^{2} + \dot{y}^{2}) + \frac{1}{2}I\dot{\theta}^{2} + \frac{1}{2}J\dot{\varphi}^{2}, \label{vdl} \non \\
\phi^{1} & = & \dot{x} - R\text{cos}\;\varphi\dot{\theta} = 0, \non \\
\phi^{2} & = & \dot{y} - R\text{sin}\;\varphi\dot{\theta} = 0, \label{vd1}
\end{eqnarray}

\noi where $m$ is the mass of the disk, $R$ is its radius, and $I,J$ are the moments of inertia about the axis perpendicular to the plane of the disk, and about the axis in the plane of the disk, respectively. 

\begin{figure}[htbp]
\begin{center}
\includegraphics[height=.3\textwidth]{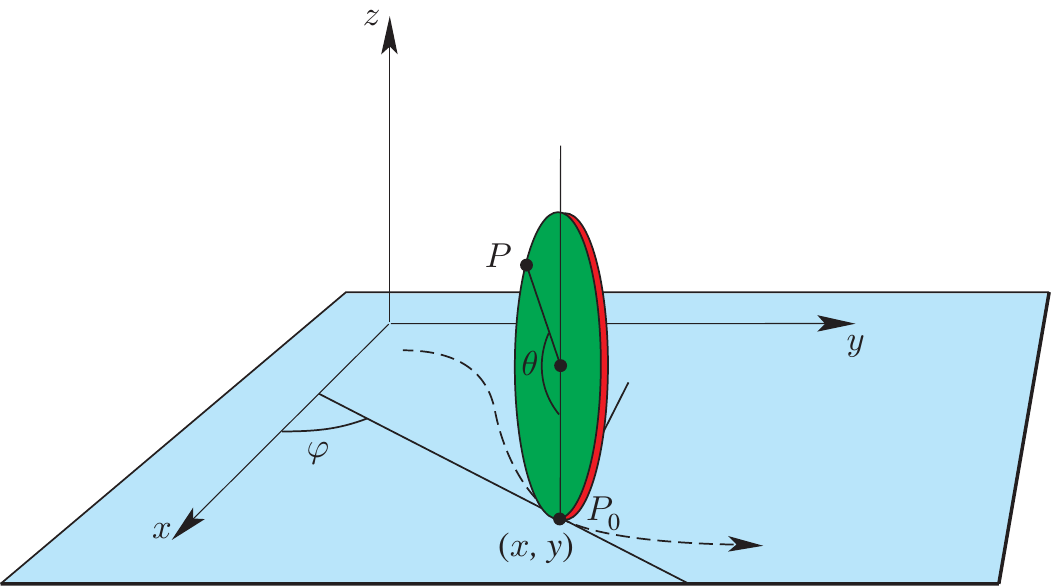}
\caption{The Vertically Rolling Disk.}
\label{vpenny2}
\end{center}
\end{figure} 

This system has an invariant measure with constant density $N$, which we can take without loss of generality to be unity. Corollary \ref{c4} applies and since $f=N=const$, then we can simply take $\dot{q}=\omega$ so that the new quasivelocities are merely the original $\dot{q}$'s. A short computation then shows that the system (\ref{vdl}) satisfies Theorem \ref{t8}, and the variational Lagrangian is computed through (\ref{lv}) to be

\beg{equation}
L_{V} (q,\dot{q})= -\frac{1}{2}m(\dot{x}^{2} + \dot{y}^{2}) + \frac{1}{2}I\dot{\theta}^{2} + \frac{1}{2}J\dot{\varphi}^{2} +mR\dot{\theta}(\dot{x}\text{cos}\;\varphi + \dot{y}\text{sin}\;\varphi). \label{vdlv}
\end{equation}

The Lagrangian (\ref{vdlv}) first appeared in \cite{FB}, and a short computation shows that applying the initial conditions $\phi^{1}(0)=0$, $\phi^{2}(0)=0$ to the Euler-Lagrange equations for $L_{V}$ reproduces the nonholonomic equations for the system (\ref{vd1}).

This simple example illustrates the case when Hamiltonization is automatic (i.e. $f=const$) and thus the system is conditionally variational as well.


\subsection{The Nonholonomic Free Particle}


Consider a nonholonomically constrained free particle with unit mass (more details can be found in \cite{B}), and Lagrangian and constraint given by

\beg{eqnarray}
&L=\frac{1}{2}\left(\dot{x}^{2} + \dot{y}^{2} + \dot{z}^{2}\right),& \non \\
&\phi(q,\dot{q}) = \dot{z} +x\dot{y}=0.& \label{nfp}
\end{eqnarray}

The system possesses an invariant measure with density $N(x)=(1+x^{2})^{-1/2}$, and thus by Corollary \ref{c4} the system is Chaplygin Hamiltonizable with $f(x)=(1+x^{2})^{-1/2}$ and quasivelocities defined by $\omega=\sqrt{1+x^{2}}\dot{r}$, where $r=(x,y)$. 

To illustrate Theorem \ref{t8}, note that ${\cal L}(q,\omega)=(1/2)f^{2}(\omega^{2}_{x}+\omega^{2}_{y}+\omega^{2}_{z})$ and that since $\widetilde{g}_{zz}=(1/2)f^{2}$, Theorem \ref{t8} applies and we have ${\cal L}_{V}$ given by:

\beg{equation}
{\cal L}_{V}(q,\omega) = \frac{1}{2(1+x^{2})}\left(\omega^{2}_{x}+\omega^{2}_{y}-\omega^{2}_{z}-2x\omega_{x}\omega_{y}\right), \non
\end{equation}

Then, computing equations (\ref{qH}) gives:

\begin{eqnarray}
\dot{\omega}_{x} &=& \frac{x\omega^{2}_{x}}{(1+x^{2})^{3/2}}, \non \\
\dot{\omega}_{y} &=& 0, \non \\
\frac{d}{dt}\left(f^{2}(\omega_{z}+x\omega_{y})\right) &=& 0. \label{npqv}
\end{eqnarray}

Now, the last line of (\ref{npqv}) reads $(d/dt)(f\phi(q,\omega))=0$, which gives the conservation law $f(q(t))\phi(q(t),\omega(t))=f(q(0))\phi(q(0),\omega(0))$. Thus, if the constraints are satisfied initially, then $\phi(q(0),\omega(0))=0$, and hence $\phi(q(t),\omega(t))=0$ for all $t$. Recalling that $\dot{q}=f\omega$, this then expresses the conservation in time of the original constraint equation (\ref{nfp}). After imposing the constraints initially, one can then use the quasivelocity definitions to then transform $\dot{\omega} \rightarrow \ddot{r}$ and recover the original nonholonomic mechanics that results from the application of the Lagrange-d'Alembert principle to the system (\ref{nfp}). Thus, although (\ref{qH}) is not Hamiltonian, as has been the theme in this paper, it is after Chaplygin's time reparameterization (and the imposition of initial conditions satisfying the constraints). Thus the nonholonomic free particle, like the vertical disk, is Hamiltonizable but since $f \neq const$ it is only conditionally variational after a reparameterization of time.  


\subsection{The Chaplygin Sphere}


The Chaplygin sphere is a sphere rolling without slipping on a horizontal plane (see \cite{B,BM}) whose center of mass is at the geometric center, but the principal moments of inertia are distinct. In Euler angles $(\theta, \varphi, \psi)$ the Lagrangian and constraints are:

\beg{eqnarray}
L = \frac{I_{1}}{2}\left(\dot{\theta}\;\text{cos}\;\varphi+\dot{\psi}\;\text{sin}\;\varphi\;\text{sin}\;\theta\right)^{2} & + & \frac{I_{2}}{2}\left(-\dot{\theta}\;\text{sin}\;\varphi+\dot{\psi}\;\text{cos}\;\varphi\;\text{sin}\;\theta\right)^{2} \non \\ & + &  \frac{I_{3}}{2}\left(\dot{\varphi}+\dot{\psi}\;\text{cos}\;\theta\right)^{2} + \frac{1}{2}\left(\dot{x}^{2}+\dot{y}^{2}\right), \label{cs1}  \non
\end{eqnarray}
\beg{eqnarray}
\phi^{1} & = & \dot{x} -\dot{\theta}\;\text{sin}\;\psi + \dot{\varphi}\;\text{cos}\;\psi\;\text{sin}\;\theta=0, \label{cs2} \non \\
\phi^{2} & = & \dot{y} +\dot{\theta}\;\text{cos}\;\psi + \dot{\varphi}\;\text{sin}\;\psi\;\text{sin}\;\theta=0. \non \label{cs3}
\end{eqnarray}

where $I_{i}$ are the moments of inertia about the center and where we have assumed the ball to have unit radius and mass. 

Since $q = (x,y,\theta, \psi, \varphi)$ and the constraints and Lagrangian are cyclic in $x,y$, we can consider this to be an abelian Chaplygin system. The system has an invariant measure whose density $N(\theta,\varphi)$ is in general non-constant \cite{BM}.

Applying Corollary \ref{hpdc} shows that there does not exist an $f$ which Hamiltonizes the three degree of freedom base dynamics given by (\ref{rechap})-(\ref{pechap}) when viewed as an abelian Chaplygin system. However, it is easily seen that $\psi$ is a nonholonomic cyclic variable and leads to the momentum conservation law $p_{\psi}=\lambda_{\psi}$. Thus we can form the constrained Routhian as in (\ref{routh}) and further reduce the dynamics to $M'=S^1 \times S^1$. We can then Hamiltonize on $M'$ through Theorem \ref{6t6}, from which (\ref{rthm}) shows that $f=N(\theta,\varphi)$. The non-canonical part of the almost-Poisson bracket (\ref{PB2}) is then computed to be

\beg{equation}
\{{\cal P'}_{1},{\cal P'}_{2}\}=-\lambda_{\psi}(I_{3}+1)f^{3}\sin\theta (I_{1}\cos^{2}\varphi+I_{2}\sin^{2}\varphi+1), \non
\end{equation}

and by the same Theorem since $dim \; M'=2$ we know that this bracket satisfies the Jacobi identity and hence is indeed a Poisson bracket. This matches the result obtained in \cite{BM} and is an example of a system that although is not Hamiltonizable at when viewed as a three degree of freedom abelian Chaplygin system is in fact Hamiltonizable on the second reduced space $M'$ of dimension 2. Moreover, it also serves to illustrate the discussion at the end of Section \ref{MCS}.


\subsection{The Snakeboard}


Another example of Theorem \ref{6t6}, whose greater importance we will discuss in the Conclusion, is the Snakeboard \cite{B,KoMa1997}. This system is modeled as a rigid body (the board) with two sets of independent actuated wheels, one on each end of the board. The human rider is modeled as a momentum wheel which sits in the middle of the board and is allowed to spin about the vertical axis, see Figure 2. 

\begin{figure}[htbp]
\begin{center}
\includegraphics[height=.3\textwidth]{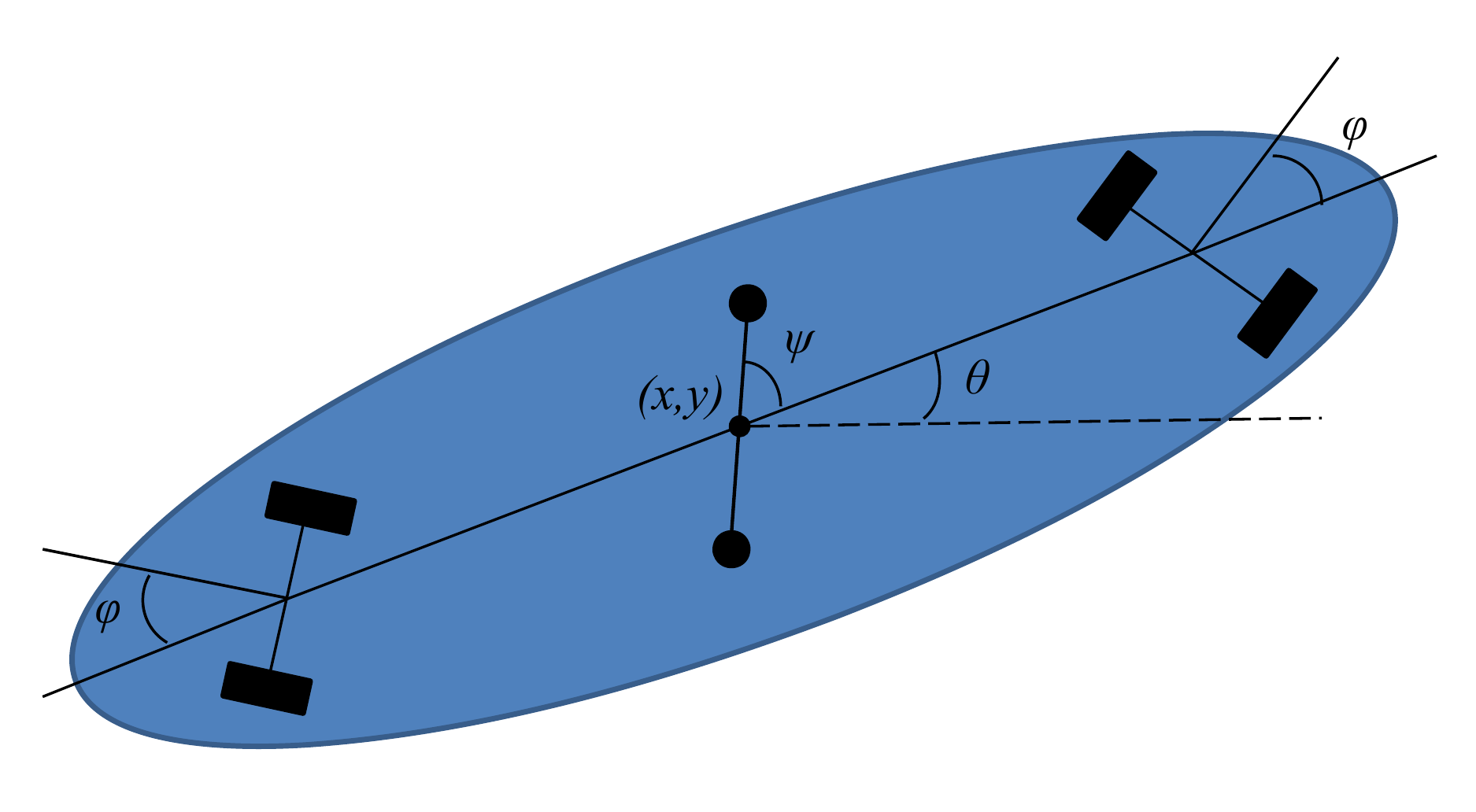}
\caption{The Snakeboard.}
\label{snake}
\end{center}
\end{figure}

The configuration space is $Q=SE(2) \times S^{1} \times S^{1}$ and the Lagrangian $L:TQ \rightarrow \mathbb{R}$ and constraints are given by:

\beg{eqnarray}\label{SnakeLandC}
&L=\displaystyle\frac{1}{2}\left(\dot{x}^{2}+\dot{y}^{2}+\dot{\theta}^{2}+\dot{\psi}^{2}+2\dot{\psi}\dot{\theta}+2\dot{\phi}^{2}\right),& \non \\
&\dot{x} = -\cot\phi\cos\theta \dot{\theta},& \non \\
&\dot{y} = -\cot\phi\sin\theta \dot{\theta},& \non
\end{eqnarray}

where we have set the mass $m$, moments of inertia, and the distance $r$ from the center of the board to its wheels equal to unity. Here $(x,y,\theta)$ represent the position and orientation of the center of the board, $\psi$ the angle of the momentum wheel relative to the board and $\phi_{1}$ and $\phi_{2}$ the angles of the back and front wheels relative to the board. Here we've made the simplification that $\phi_{1}=-\phi_{2}$, as in \cite{B,KoMa1997}.

As stated, we can view this system as an abelian Chaplygin nonholonomic system with three degrees of freedom. Its equations of motion are given in \cite{KoMa1997} as:

\begin{center}
$\begin{array}{ll}
\dot{p}_{\theta}=-\displaystyle\frac{1}{2}\sec\phi\csc\phi(p_{\theta}-p_{\psi})p_{\phi}, & \dot{\theta}=\tan^{2}\phi(p_{\theta}-p_{\psi}), \\
\dot{p}_{\phi}=0, & \dot{\phi}=\displaystyle\frac{1}{2}p_{\phi}, \\
\dot{p}_{\psi}=0, & \dot{\psi} = \displaystyle\frac{p_{\psi}-\sin^{2}\phi p_{\theta}}{\cos^{2}\phi}.
\end{array}$
\end{center}

Since this system satisfies the conditions of Theorem \ref{6t6} we can set $p_{\psi}=\lambda_{\psi}=const.$ and focus on Hamiltonizing the reduced system. Given that this reduced system has the invariant measure $N(\phi)=\tan\phi$, which is independent of $\psi$, by Corollary \ref{c4} $f=N$. The non-canonical part of the almost-Poisson bracket (\ref{PB2}) is then computed to be:

\beg{equation}
\{{\cal P}'_{1},{\cal P}'_{2}\}=\sec^{2}\phi\lambda_{\psi}, \non
\end{equation} 

and by the same Theorem we know that this bracket satisfies the Jacobi identity (since the reduced system has two degrees of freedom) and is thus a Poisson bracket. 


\subsection{The Chaplygin Sleigh}\label{CSS}


The Chaplygin Sleigh \cite{B, B2,BM5,Chap, Chap2, NF} consists of a rigid body in the plane which is supported at three points, two of which slide freely without friction while the third is a knife edge, a constraint that allows no motion perpendicular to its edge. The configuration manifold $Q=\mathbb{R}^{2} \times S^{1}$, where $(x,y)$ are the coordinates of the contact point while $\theta$ is the angle the knife edge makes with the $x$-axis, see Figure 3 below. Moreover, we suppose here that the center of mass of the system $C$ is not on top of the knife edge (if it is, then one can show \cite{B} that the sleigh reduces to another nonholonomic system known as the \itl{knife edge}, which possesses an invariant measure).  

\begin{figure}[htbp]
\begin{center}
\includegraphics[height=.3\textwidth]{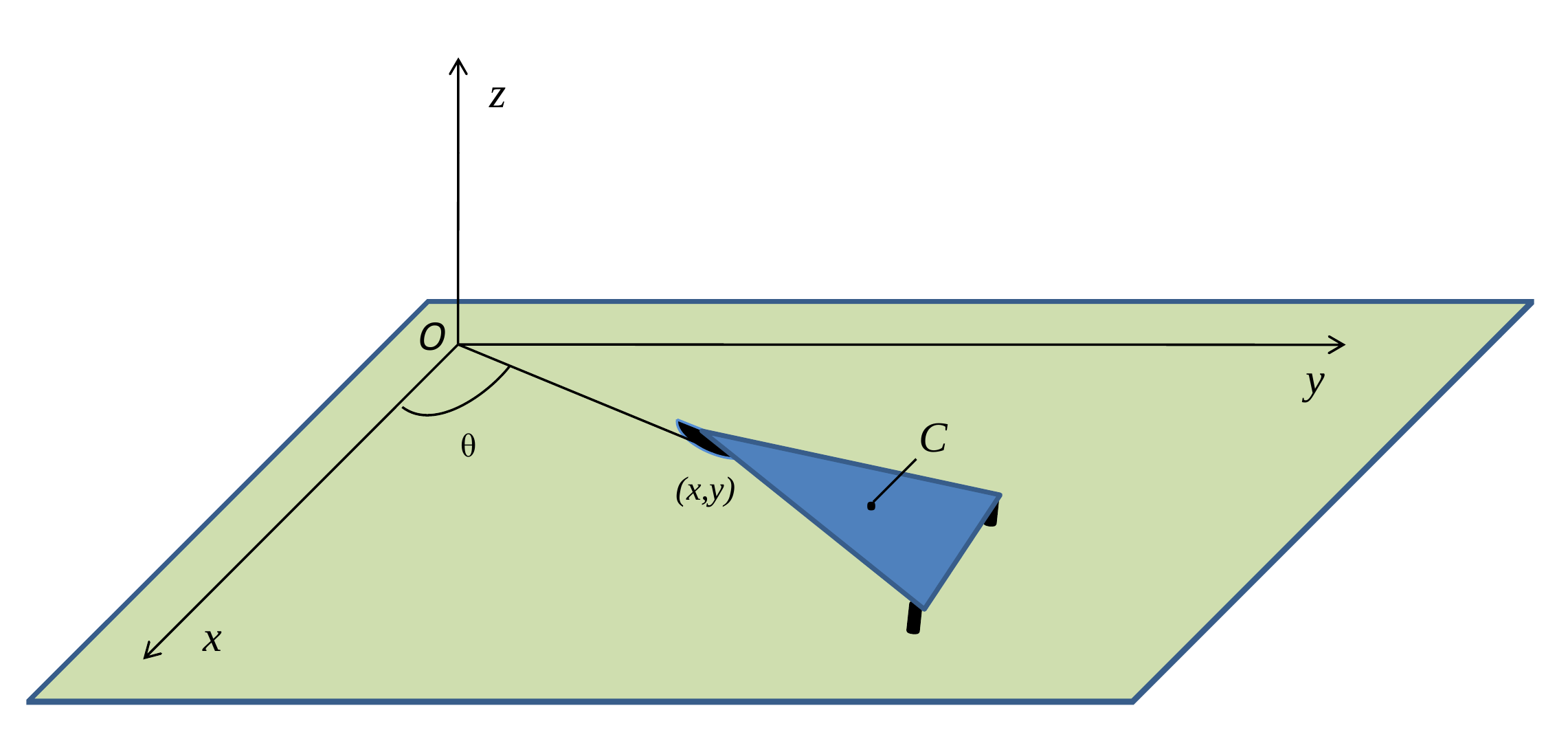}
\caption{The Chaplygin Sleigh.}
\label{chapsl}
\end{center}
\end{figure} 
\noi The Lagrangian $L$ and constraints are given by:

\beg{eqnarray}
&L = \displaystyle\frac{1}{2}\left(\dot{x}^{2}+\dot{y}^{2}+2\dot{\theta}^{2}-2(\dot{x}\sin\theta + \dot{y}\cos\theta)\dot{\theta}\right),& \label{csl} \non \\
&\dot{y}\cos\theta - \dot{x}\sin\theta = 0,& \label{csc} \non 
\end{eqnarray}

where for simplicity we have set all parameters to unity. Since the Lagrangian and constraint are left invariant on the Lie group $G=SE(2)$ we can treat the problem within the Euler-Poincar\'e-Suslov framework. Defining $\xi=g^{-1}\dot{g}$, where $g=(x,y,\theta)$, we can write the Lagrangian $L$ in terms of $\xi$ as $l(\xi)=\xi_{3}^{2}+(1/2)(\xi_{1}^{2}+\xi_{2}^{2})+\xi_{2}\xi_{3}$, and the constraint as $\xi^{2}=0$.  

With the structure constants given by $C^{2}_{13}=-1=-C^{1}_{23}$ and all other zero we see that $f=const.$ satisfies Corollary \ref{hpdc}, which agrees with the recent result of \cite{Nar2}.

This system is of critical importance in the study of Hamiltonization since unlike Proposition \ref{6p4}, the Chaplygin sleigh shows that just because a system is Hamiltonizable does not imply that it possesses an invariant measure. Indeed, although it is well-known the Chaplygin sleigh does not possess an invariant measure \cite{B,Nar2}, as we've seen above this system is nonetheless Hamiltonizable.  Thus, unlike for the nonabelian Chaplygin case, the Hamiltonizability of Euler-Poincar\'e-Suslov systems does not automatically imply that the system possesses an invariant measure, and thus Chaplygin's Reducibility Theorem becomes inapplicable (due to the non-existence of an invariant measure)\footnote{However, it is interesting to note that Chaplygin {\em did} apply his reducing multiplier method to the Chaplygin sleigh in \cite{Chap2}, but only after introducing ``quasicoordinates.''}. However, thanks to the results of Corollary \ref{hpdc}, we may still be able to Hamiltonize, or, more properly, ``Poissonize'' (which would be the better term here since there isn't an invariant measure, as discussed in the Introduction). 


\subsection{A Mathematical Example}


Consider the following mathematical example due to Iliyev \cite{Il}. The Lagrangian and constraints are given by:

\beg{eqnarray}
L & = & \frac{1}{2}\left((\dot{q}^{1})^{2}+(\dot{q}^{2})^{2}+(\dot{q}^{3})^{2}+(\dot{q}^{4})^{2}+(\dot{q}^{5})^{2} \right), \non \\
\dot{q}^{4} & = & \dot{q}^{2}\tan (q^{1}),  \non \\
\dot{q}^{5} & = & \dot{q}^{3}\tan (q^{1}). \non
\end{eqnarray}
This is a nonholonomic system with three degrees of freedom ($m=3$), and thus Chaplygin's Theorem is inapplicable. Within our framework, it can most easily be treated as an abelian Chaplygin system. Solving the conditions in (\ref{condhdf}) in MAPLE yields $f=\cos (q^{1})$. Moreover, as a check of Proposition \ref{6p4}, one can show that the system's invariant measure density is $N=\cos^{2}(q^{1})$, which indeed is equal to $f^{m-1}$, as the Proposition suggests. 


\section{Conclusion and Future Directions}


Chaplygin's Reducing Multiplier Theorem has long allowed an interesting investigation of some nonholonomic systems in terms of the quasi-periodic orbits that result from the consideration of the time reparameterization $d\tau = f dt$ and the Hamiltonian-like structure it produces. Perhaps because of this, and its success in studying nonholonomic systems using methods from unconstrained mechanics, it has attracted much attention in the recent decades as interest in nonholonomic systems has grown. However, as we have mentioned, the use of Chaplygin's results, and much of the subsequent research that has followed it, has been confined to systems possessing an invariant measure and, typically, also in two degrees of freedom. In addition, and partly due to these confines, Hamiltonizability of nonholonomic systems with symmetry in arbitrary dimensions has remained untouched (with the only results \cite{FJ3} and \cite{Jov2} known to the authors arising by construction). 

The present work addresses these two main aspects of Chaplygin's work, extending the results to nonholonomic systems in arbitrary degrees of freedom not necessarily possessing an invariant measure (a central assumption of the research on Chaplygin's work to date). As such, this latter result alone represents a possibly new direction for the study of the integrability of nonholonomic systems, with a Hamilton-Jacobi theory based on it now possible (it would be interesting to develop this and compare it to the Hamilton-Jacobi theory of nonholonomic systems recently presented in \cite{Ponte}. In fact, as we have mentioned, Chaplygin's Theorem was used in conjunction with the Hamilton-Jacobi method implicitly in \cite{Chap2,NF}, albeit in ``quasicoordinates.''). The local conditions (\ref{c1})-(\ref{c5}), or their special cases (\ref{condhdf}) and (\ref{epslocal}), also enable the search for Hamiltonizable nonholonomic systems to be converted into the search for solutions to certain partial differential equations, a task which can be considerably simplified by making use of any of today's mathematical software packages and which eliminates the guesswork involved in current uses of Chaplygin's Theorem.  

We should also point out the interesting role that symmetry plays in Chaplygin Hamiltonization. For example, considering the Chaplygin sleigh as an abelian Chaplygin system it is immediately seen to be impossible to Chaplygin Hamiltonize (the equations (\ref{condhdf}) have no solution), yet as we showed in Section 4.5 it {\em is} Chaplygin Hamiltonizable when considered as an Euler-Poincar\'e-Suslov system. This suggests to us that the choice of symmetry group affects the Hamiltonizability of the system in question. We expect to pursue these issues in future research.

Finally, we note that the multi-dimensional Veselova system and multi-dimensional Chaplygin sphere have recently been Hamiltonized in \cite{FJ3} and \cite{Jov2}, respectively. However, the methods and conditions for Hamiltonization presented here are inapplicable to those Hamiltonizations due to the particular Hamiltonization methods used by the authors. 

In the former, the authors constructed redundant coordinates and showed that the solutions of the multi-dimensional Veselova system can be mapped isomorphically into the solutions of an associated \itl{different} Hamiltonian system known as the Neumann system. Within the framework of the methods presented here and in our previous research \cite{BFM}, this would be equivalent to the statement that after an appropriate time reparameterization, applying the inverse problem of the calculus of variations to the resulting system would yield the Neumann Lagrangian as a solution. 

In the latter case, the author Hamiltonizes the multi-dimensional Chaplygin sphere by constructing redundant coordinates and effecting a time-reparameterization. It is then shown that the reduced mechanics of the higher dimensional nonholonomic Chaplygin sphere emerge as the restriction to the invariant submanifolds of the Hamiltonian system resulting from the time reparameterization. In our previous research \cite{BFM} we called these type of systems \itl{associated second-order systems}. However, the main difference between our work there and the construction in \cite{Jov2} is that we constructed associated second-order systems for the original nonholonomic system (not the time reparameterized one). 

Given the above discussion, we therefore expect that the aforementioned multi-dimensional Hamiltonizations can be realized as special cases of a synthesis of the general (yet mostly disjoint) methods presented here and in earlier work \cite{BFM} (see also \cite{FT}).


\section{Acknowledgments}


The research of OEF and AMB was supported in part by the Rackham
Graduate School of the University of Michigan, through the Rackham
Science award and the AGEP Fellowship, and through NSF grants DMS-0604307 and DMS-0907949 respectively.
TM acknowledges a Marie Curie Fellowship within the 6th European Community Framework Programme and a Postdoctoral Fellowship of the Research Foundation - Flanders (FWO). We would also like to thank the reviewer for many useful comments.


\section{Appendix}


\subsection{The Components of the Quasivelocity AP Bracket}


The components of (\ref{1})-(\ref{3}) are given by:

\beg{eqnarray}
\hat{A}^{k}_{ij} & = & fC^{a}_{bd}e^{b}_{i}e^{d}_{j}M_{a\gamma}G^{\gamma\alpha}G^{k}_{\alpha}-\left(fK^{k}_{ji}-\overline{C}^{k}_{ij}\right), \label{abe} \\
\hat{B}^{\alpha}_{ij} & = & -fC^{a}_{cd}e^{c}_{i}e^{d}_{j}M_{a\gamma}G^{\gamma\alpha}, \label{bbe} \\
\hat{C}^{k}_{i \alpha} & = & \delta^{k}_{i}\left(\frac{\partial f}{\partial r^{\alpha}}-\frac{\partial f}{\partial g^{\sigma}}g^{\sigma}_{d}A^{d}_{\alpha}\right)+fF^{a}_{i \alpha}\left(g_{ab}\Gamma^{bk}-M_{a\gamma}G^{\gamma\beta}G^{k}_{\beta}\right), \label{cbe} \\
\hat{D}^{\beta}_{i \alpha} & = & fF^{a}_{i \alpha}M_{a\gamma}G^{\gamma\beta}-\frac{\partial f}{\partial g^{\sigma}}g^{\sigma}_{d}e^{d}_{i}\delta^{\beta}_{\alpha}, \label{dbe} \\
\hat{E}^{k}_{\alpha\beta} & = & f{\cal B}^{b}_{\alpha\beta}\left(M_{b\gamma}G^{\gamma\epsilon}G^{k}_{\epsilon}-g_{bd}\Gamma^{dk}\right), \label{ebe} \\
\hat{F}^{\gamma}_{\alpha\beta} & = & \frac{\partial f}{\partial g^{\sigma}}g^{\sigma}_{b}\left({\cal A}^{b}_{\alpha}\delta^{\gamma}_{\beta}-{\cal A}^{b}_{\beta}\delta^{\gamma}_{\alpha}\right)+\left(fK^{\gamma}_{\beta\alpha}-C^{\gamma}_{\alpha\beta}\right), \label{fbe}
\end{eqnarray}
where 

\beg{eqnarray}
K^{\gamma}_{\beta\alpha} & = & M_{b\epsilon}G^{\epsilon\gamma}{\cal B}^{b}_{\beta\alpha}, \label{kchap} \\
K^{k}_{ji} & = & g_{ab}C^{a}_{cd}e^{c}_{i}e^{d}_{j}\Gamma^{bk}, \label{keps} \\
C^{\gamma}_{\alpha\beta} & = & \delta^{\gamma}_{\beta}\frac{\partial f}{\partial r^{\alpha}}-\delta^{\gamma}_{\alpha}\frac{\partial f}{\partial r^{\beta}}, \label{cchap} \\
\overline{C}^{k}_{ij} & = & \frac{\partial f}{\partial g^{\sigma}}g^{\sigma}_{d}\left(e^{d}_{j}\delta^{k}_{i}-e^{d}_{i}\delta^{k}_{j}\right), \label{ceps} 
\end{eqnarray}
with $M_{a\alpha}=g_{a\alpha}-g_{ab}{\cal A}^{b}_{\alpha}$.


\subsection{Calculation of the Jacobi Identity for the Quasivelocity AP Bracket}


Here we illustrate the calculation of the Jacobi identity for the bracket (\ref{qvapb}). Since it is well-known \cite{B} that the Jacobi identity is satisfied iff it is satisfied for the component functions, we need only calculate it for all combinations of $x=(r^{\gamma},{\cal P}_{i},{\cal P}_{\alpha})$, i.e. we require $\{x^{I},\{x^{J},x^{K}\}'_{\overline{M}}\}'_{\overline{M}} + \text{cyclic}=0$ for all $I,J,K=(a,i,\alpha)$. 

As an example, consider $x=(r^{\gamma},{\cal P}_{\alpha},{\cal P}_{\beta})$. Then we have:

\beg{eqnarray}
&\{r^{\gamma},\{{\cal P}_{\alpha},{\cal P}_{\beta}\}'_{\overline{M}}\}'_{\overline{M}} + \text{cylic} =0,& \non \\
& \implies \{r^{\gamma},\frac{1}{f}\hat{F}^{\epsilon}_{\alpha\beta}{\cal P}_{\epsilon}\}'_{\overline{M}}=0,& \non \\
&\implies \frac{1}{f^{2}}\delta^{\gamma}_{\epsilon}\hat{F}^{\epsilon}_{\alpha\beta}=0, \non
\end{eqnarray}
which gives the third equation in (\ref{c1}). Similarly, considering $x=(r^{\gamma},{\cal P}_{i},{\cal P}_{\beta})$ gives:

\beg{eqnarray}
&\{r^{\gamma},\{{\cal P}_{i},{\cal P}_{\beta}\}'_{\overline{M}}\}'_{\overline{M}} + \text{cylic} =0,& \non \\
& \implies \{r^{\gamma},\frac{1}{f}\hat{D}^{\epsilon}_{i \beta}{\cal P}_{\epsilon}\}'_{\overline{M}}=0,& \non \\
&\implies \frac{1}{f^{2}}\delta^{\gamma}_{\epsilon}\hat{D}^{\epsilon}_{i \beta}=0, \non
\end{eqnarray}
which gives the second equation in (\ref{c1}). Similar computations lead to the remaining conditions in (\ref{c1}).



\begin{thebibliography}{99}


\bibitem{Ap} Appell, P. Sur des transformations de movements, {\em J. reine und angew. Math.}, 110 (1892), 37-41.

\bibitem{B} Bloch, A.M. {\em Nonholonomic Mechanics and Control}, Springer NY (2003).

\bibitem{B2} Bloch, A.M. Asymptotic Hamiltonian dynamics: the Toda lattice, the three wave interaction, and the nonholonomic Chaplygin sleigh, {\em Physica D}, 141 (2000), 297-315.

\bibitem{BFM} Bloch, A.M., Fernandez, O.E. and Mestdag, T. Hamiltonization of Nonholonomic Systems and the Inverse Problem of the Calculus of Variations, {\em Rep. Math. Phys.}, 63 (2009), 225-249. 

\bibitem{BKMM} Bloch, A.M., Krishnaprasad, P.S., Marsden, J.E. and Murray, R. Nonholonomic mechanical systems with symmetry, {\em Arch Rat. Mech. An.}, 136 (1996), 21-99.

\bibitem{BMZ} Bloch, A.M., Marsden, J.E. and Zenkov D.V. Quasivelocities and Symmetries in Nonholonomic Systems, {\em Dynamical Systems}, 24 (2009), 187-222.

\bibitem{BM} Borisov, A.V. and Mamaev, I.S. Isomorphism and Hamilton representation of some nonholonomic systems, {\em Siberian Math. J.}, 48(1) (2007), 26-36.

\bibitem{BM2}\indent --- Conservation Laws, Hierarchy of Dynamics and Explicit Integration of Nonholonomic Systems, {\em Reg. Chaotic Dyn.}, 13(5) (2008), 443-490.

\bibitem{BM3}\indent --- The rolling motion of a rigid body on a plane and a sphere. Hierarchy of dynamics, {\em Regul. Chaotic Dyn.}, 7(2) (2002), 177-200.

\bibitem{BM4}\indent --- Strange Attractors in Rattleback Dynamics, {\em Physics-Uspekhi}, 46(4) (2003,) 393-403.

\bibitem{BM5}\indent --- The dynamics of a Chaplygin sleigh, {\em J. of Applied Math. and Mech.},  73(2) (2009), 156-161. 

\bibitem{BMK} Borisov, A.V., Mamaev, I.S. and Kilin, A.A. The rolling motion of a ball on a surface. New integrals and hierarchy of dynamics, {\em Regul. Chaotic Dyn.}, 7(2) (2002), 201-219.

\bibitem{CCD} Cantrijn, F., Cortes, J., de Leon, M. and Martin de Diego. On the geometry of generalized Chaplygin systems, {\em Math. Proc. Cambridge Philos. Soc.}, 132(2) (2002), 323-351.

\bibitem{Chap} Chaplygin, S.A. On a rolling sphere on a horizontal plane, {\em Mat. Sbornik}, 24 (1903), 139-168 (in Russian) and {\em Reg. Chaotic Dyn.}, 7(2) (2002), 131-148 (in English).

\bibitem{Chap2}\indent --- On the theory of motion of nonholonomic systems. Theorem on the reducing multiplier, {\em Mat. Sbornik}, 28(2) (1911), 303-314 (in Russian) and {\em Reg. Chaotic Dyn.}, 13(4) (2008), 369-376 (in English).

\bibitem{Chap3}\indent --- On the theory of motion of nonholonomic systems; examples of application of the reducing multiplier method. {\em Collection of works, vol. III. M.-L.: Gostechizdat.}, (1950), 248-259.

\bibitem{C} Cortes, J. {\em Geometric, control and numerical aspects of nonholonomic systems}, Springer NY (2002).

\bibitem{E} Ehlers, K., Koiller, J., Montgomery, R. and Rios, P.M. Nonholonomic systems via moving frames: Cartan equivalence and Chaplygin Hamiltonization, In: {\em The breadth of symplectic and 
Poisson geometry}, Progr.\ Math.\ 232, Boston (2005), 75--120. 

\bibitem{FJ} Fedorov, Y.N. and Jovanovi\'c, B. Quasi-Chaplygin Systems and Nonholonomic Rigid Body Dynamics, {\em Lett. Math. Phys.}, 76(2-3) (2006), 215-230.

\bibitem{FJ2}\indent --- Integrable Nonholonomic Geodesic Flows on Compact Lie Groups, {\em Top. Methods in the theory of Integrable Systems}, (2005), 115-152.

\bibitem{FJ3}\indent --- Nonholonomic LR Systems as Generalized Chaplygin systems with an Invariant Measure and Geodesic Flows on Homogeneous Spaces, {\em J. Non. Sci.}, 14 (2004), 341-381.

\bibitem{FT} Fernandez, O.E. {\em The Hamiltonization of Nonholonomic Systems and its Applications}, Ph.D. Thesis (2009), The University of Michigan. 

\bibitem{FB} Fernandez, O.E. and Bloch, A.M. Equivalence of the Dynamics of Nonholonomic and Variational Nonholonomic Systems for certain Initial Data, {\em J. Phys. A: Math. Theor.}, 41 (2008).

\bibitem{Nar1} Garcia-Naranjo, L.C. Reduction of Almost Poisson brackets and Hamiltonization of the Chaplygin Sphere, {\em eprint arXiv:0808.0854}.

\bibitem{Nar2}\indent --- Reduction of Almost Poisson Brackets for Nonholonomic Systems on Lie Groups, {\em Reg. Chaotic Dyn.}, 12(4) (2007), 365-388.  

\bibitem{Ponte} Iglesias-Ponte, D., de Le\'on, M. and Martin de Diego, D. Towards a Hamilton-Jacobi theory for nonholonomic mechanical systems, {\em J. Phys. A: Math. Theor.}, 41 (2008), 015205.

\bibitem{Il} Iliyev, IL. On the conditions for the existence of the reducing chaplygin factor, {\em Prikl. Mat. Mehk.}, 49(2) (1985), 295-301.

\bibitem{Jov2} Jovanovi\'c, B. Hamiltonization and integrability of the Chaplygin sphere in $\mathbb{R}^{n}$, {\em eprint arXiv:0902.4397}.

\bibitem{Ker} Kharlamova, E.I. Rigid Body motion about a fixed point under nonholonomic constraint, {\em Proc. Donetsk. Industr. Inst.}, 20(1) (1957), 69-75.

\bibitem{Koi} Koiller, J. Reduction of some classical non-holonomic systems with symmetry, {\em Arch. Rational Mech. Anal.}, 118 (1992), 113-148.

\bibitem{KoMa1997}
   Koon, W.S. and Marsden, J.E.
   The {H}amiltonian and {L}agrangian approaches to the dynamics of
   nonholonomic systems, {\em  Reports on Math Phys. }, 40 (1997),
   21-62.

\bibitem{MR} Marsden, J.E. and Ratiu, T.S. {\em Introduction to Mechanics and Symmetry}, Second Ed., Springer NY (1999).

\bibitem{CMR2} Marsden, J.E. and Scheurle, J. The reduced Euler-Lagrange equations, {\em Fields Inst. Com.}, 1 (1993), 139-164.

\bibitem{NF} Neimark, J.I. and Fufaev, N.A. {\em Dynamics of Nonholonomic Systems}, Amer.\ Math.\ Soc.\ (1972).

\bibitem{Ram} Ramos, A. Poisson structures for reduced non-holonomic systems, {\em J. Phys. A}, 37(17) (2004), 4821-4842. 

\bibitem{Stan} Stanchenko, S.V. Nonholonomic Chaplygin Systems, {\em Prikl. Mat. Mehk.}, 53(1) (1989), 16-23. 

\bibitem{Sum} Sumbatov, A.S. Nonholonomic Systems, {\em Reg. Chaotic Dyn.}, 7(2) (2002), 221-238.

\end{thebibliography}
\end{document}